\documentclass[
 reprint,
 amsmath,amssymb,
 aps,
]{revtex4-2}
\usepackage{graphicx}
\usepackage{dcolumn}
\usepackage{bbm}
\usepackage{mathtools}
\usepackage{amsthm}
\usepackage{booktabs} 
\usepackage{comment}

\newtheorem{theorem}{Theorem}
\newcommand{\longthmtitle}[1]{%
  \leavevmode\par\noindent
  {\normalfont\bfseries #1\par}%
  \smallskip\noindent
}
\newtheorem{definition}{Definition}
\newtheorem{lemma}[theorem]{Lemma}
\newtheorem{corollary}[theorem]{Corollary}

\usepackage{adjustbox}
\usepackage{tikz}
\usetikzlibrary{quantikz2}

\begin{document}


\title{Quantum Channel Polynomial Processing}
\author{Tianhan Liu}
 \email{tianhan.liu@iqm.tech}
\affiliation{%
IQM Quantum Computers, Georg-Brauchle-Ring 23-25, 80992 Munich, Germany}%
\author{Fedor \v{S}imkovic IV}
\affiliation{%
IQM Quantum Computers, Georg-Brauchle-Ring 23-25, 80992 Munich, Germany}%
\author{Martin Leib}
\affiliation{%
IQM Quantum Computers, Georg-Brauchle-Ring 23-25, 80992 Munich, Germany}%


\date{\today}

\begin{abstract}
We introduce a quantum algorithmic framework based  on probabilistic mixtures of unitary channels that, similar to the framework of quantum singular value transformations, enables the application of arbitrary polynomials of Hermitian operators onto arbitrary initial states. We show that our framework supports a flexible tradeoff between sample- and query complexity ranging from optimal query complexity, meaning linear in evolved time,  and exponentially scaling sample complexity to quadratic in evolved time with polynomial sample complexity, for real and imaginary time evolution. In both cases we are able to prove an approximation error converging exponentially as a function of the query complexity. Combined with the considerably lower quantum circuit complexity, compared to quantum singular value transformations with a linear combination of unitaries block encoding, we argue that our framework can be seamlessly scaled from NISQ to fault-tolerant quantum computing.   


\end{abstract}

\maketitle

\section{Introduction}

Quantum algorithms are expected to provide asymptotic advantages over classical counterparts for a range of problems, including quantum search~\cite{PhysRevLett.113.210501, PhysRevLett.95.150501}, factoring~\cite{Shor_1997}, quantum simulation~\cite{Berry_2015, Low2019hamiltonian}, and differential equations~\cite{Harrow_2009}. A recurring primitive in many of these algorithms is the implementation of functions of an operator, and in particular functions of a Hamiltonian, $f(H)$, within a quantum process~\cite{mrtc_unification_21}. Polynomial approximation provides a natural route to this task: a target function $f$ is approximated by a polynomial of degree $d$, after which the problem becomes one of realizing the corresponding polynomial transformation of the Hamiltonian $H$ on quantum hardware.

Quantum singular value transformation (QSVT) provides a powerful coherent realization of this idea. Given access to a suitable unitary encoding of $H$, QSVT implements polynomial transformations of the encoded operator using $\mathcal{O}(d)$ applications of the underlying unitary, where $d$ is the degree of the approximating polynomial. When such an encoding is available, this scaling makes QSVT an asymptotically optimal framework in the query model. Together with the theory of polynomial approximation and interpolation~\cite{10.5555/3384673, 10.5555/2024608}, QSVT and related block-encoding techniques provide a central theoretical framework for Hamiltonian simulation~\cite{Low2019hamiltonian}, ground-state preparation~\cite{PRXQuantum.3.040305}, thermal-state preparation~\cite{chen2023efficientexactnoncommutativequantum}, matrix arithmetic~\cite{Gily_n_2019}, and other Hamiltonian-function algorithms.

The practical use of QSVT is limited by the need to encode the Hamiltonian of interest into a unitary operation. In conventional block-encoding approaches, such as linear combination of unitaries (LCU)~\cite{Low2019hamiltonian, 2012, low2019hamiltoniansimulationinteractionpicture} or sparse-matrix encodings~\cite{camps2023explicitquantumcircuitsblock}, the coefficients of Hamiltonian terms or sparse matrix entries are loaded into ancillary registers, often using quantum read-only memory (QROAM)~\cite{Low_2024}, while controlled operations select the corresponding unitary terms. For quantum chemistry Hamiltonians, where the number of terms typically scales as $\mathcal{O}(N^4)$, such coherent selection procedures imply substantial ancillary overhead, many controlled operations, and large Toffoli counts. As a result, the standard block-encoded route to QSVT is most naturally suited to large-scale fault-tolerant quantum computers.

Several approaches seek to reduce this overhead while remaining within the block-encoding framework. Rather than coherently selecting from a generic list of Hamiltonian terms, these methods exploit additional structure of the operator to simplify state preparation, control logic, or the encoded representation itself. Examples include variational block-encoding methods~\cite{Kikuchi_2023}, tensor-network encodings of matrix product operators~\cite{Nibbi_2024}, Dicke-state-based constructions~\cite{dellachiara2025efficientlcublockencodings}, and chemistry-specific encodings~\cite{simon2025ladderoperatorblockencoding, liu2024efficientquantumcircuitblock, PhysRevX.8.041015, PRXQuantum.2.030305}. While these approaches can lower particular resource costs, they still require a coherent unitary encoding of the Hamiltonian, and their efficiency depends on optimization, compressibility, or problem-specific algebraic structure.

A distinct approach is to avoid coherent Hamiltonian encoding and represent the Hamiltonian statistically through sampled circuits. Product formulas, or Trotterization~\cite{PhysRevLett.79.2586, 1985JMP....26..601S}, implement Hamiltonian dynamics by applying gates generated by individual Hamiltonian terms, with circuit complexity depending on the target precision~\cite{Childs_2021} and the number of Hamiltonian terms. Randomized product formulas such as QDrift~\cite{Qdrift,nakaji2023qswift,lee2026qshift} replace deterministic traversal over Hamiltonian terms by probabilistic sampling, thereby trading part of the circuit depth for repetitions; in suitable regimes, the sampling cost can be controlled using concentration properties of martingale processes~\cite{PRXQuantum.2.040305}. Randomized LCU methods~\cite{Zeng_2025, sun2025randomisedcompositelinearcombinationofunitariesrole, wang2025randomizedquantumsingularvalue} apply a related sampling principle to conditional Pauli circuits. These methods show that coherent Hamiltonian selection can be replaced by stochastic circuit ensembles in certain settings. However, they are primarily designed for Hamiltonian time evolution or sampled LCU constructions, and do not directly provide a general polynomial-transformation framework for arbitrary functions, $f(H)$, and they often entail an exponential sampling overhead.

This leaves an algorithmic gap between block-encoded polynomial transformations and stochastic Hamiltonian simulation. QSVT provides a general polynomial framework but relies on coherent unitary block encodings of the Hamiltonian. Product-formula and randomized simulation methods reduce coherent circuit requirements by using sampled circuits, but their structure is not naturally adapted to arbitrary Hamiltonian functions. The question is therefore whether polynomial Hamiltonian-function approximation can be combined with stochastic circuit sampling, without constructing a deterministic coherent block encoding of $H$.

In this work, we introduce a stochastic quantum algorithm framework called quantum channel polynomial processing (QCPP) for implementing Hamiltonian functions in quantum channels. The construction follows the polynomial-approximation perspective underlying QSP, while replacing coherent block encoding by stochastic encodings of Hamiltonian commutators and anti-commutators at the channel level. These encodings are then used to formulate a corresponding signal-processing procedure for target functions of Hamiltonians. For a degree-$d$ interpolating polynomial, each sampled circuit contains at most $2d$ conditional Pauli rotations, while additional Hamiltonian terms enter through the sampling distribution rather than through coherent selection hardware.

This formulation shifts the resource requirements from coherent Hamiltonian selection to stochastic sampling. Each circuit instance is substantially simpler than a conventional block-encoded QSP circuit and avoids many of the coherent-control requirements that dominate standard implementations~\cite{Low_2016}, while sample complexity becomes the relevant additional resource. Thus, our proposed QCPP method occupies a different point in the algorithmic design space: it retains the use of polynomial approximations to target Hamiltonian functions while trading coherent circuit complexity for repetitions. We study the sample complexity of this stochastic compilation procedure for real- and imaginary-time  evolution as a low-resource application. The resulting circuit structures indicate a route toward implementing QSVT-inspired Hamiltonian-function methods on near- and intermediate-term quantum computers.

The paper is structured as follows: We start in Section \ref{sec:algorithm_framework} by introducing the general stochastic quantum algorithm framework of QCPP, explaining along the way how it implements polynomials of Hermitian operators. In Section \ref{sec:sample_query} we prove fundamental relationships illustrating the sample- and query complexity tradeoffs of the method. We finish the article with general remarks in Section \ref{sec:general_remarks} relating to commutativity and the relation of QCPP to QSVT and provide concluding thoughts in Section \ref{sec:conclusion}. 

\begin{figure}[ht]
    \centering    \includegraphics[width=0.92\linewidth]{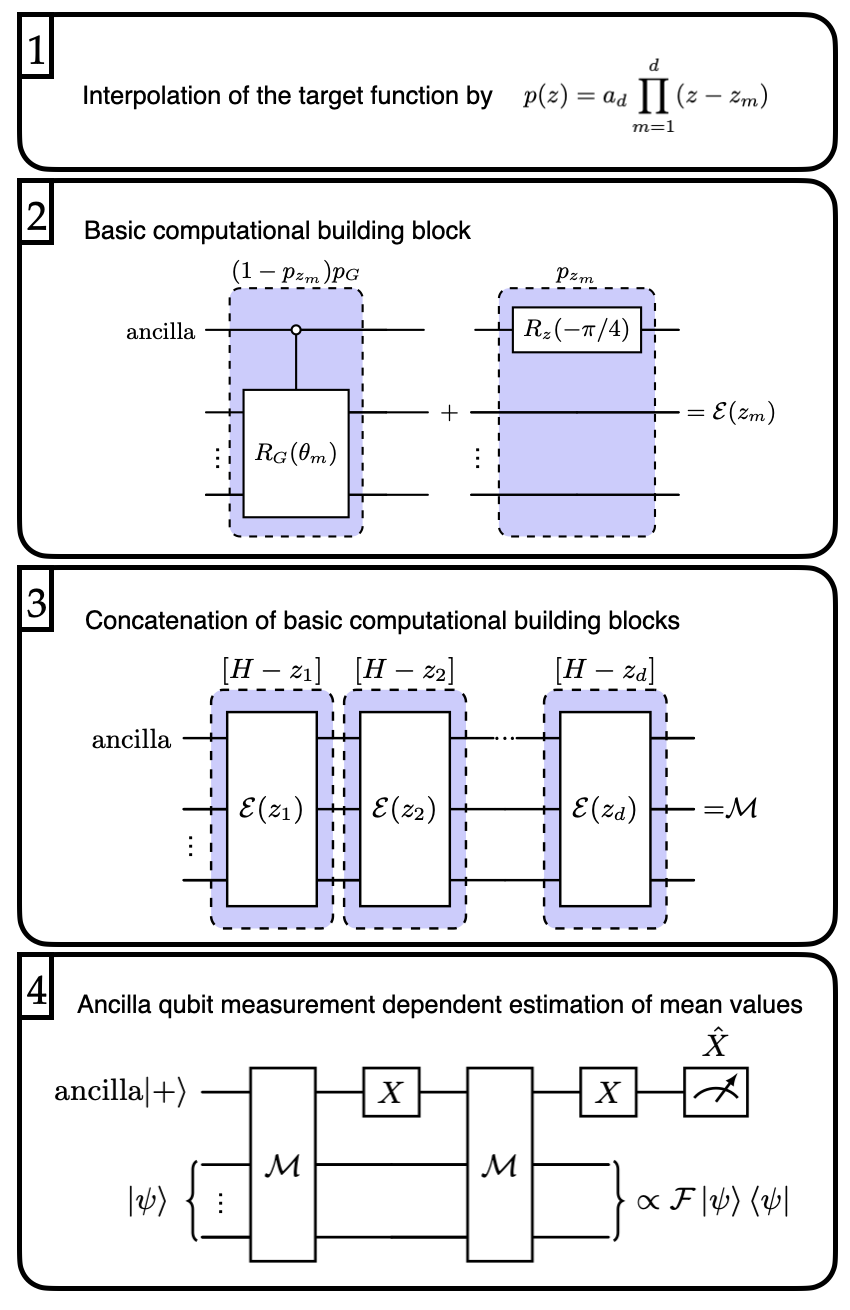}
    \caption{
Workflow of Quantum Channel Polynomial Processing
(QCPP).
(1) The target function $f$ is approximated by
a degree-$d$ polynomial $p$.
(2) Each root of $p$ defines a basic computational
building block, implemented by stochastic sampling
of unitary circuits.
(3) Concatenating the $d$ building blocks
constructs the quantum channel $\mathcal{M}$.
(4) The channel $\mathcal{M}$ is applied twice,
with Pauli-$X$ gates acting on the ancilla as shown.
The ancilla measurement outcome is used to
assign signs to the measured observables,
enabling statistical reconstruction of the
desired polynomial transformation.
}
    \label{fig:workflow_qcpp}

\end{figure}

\section{Algorithm Framework}\label{sec:algorithm_framework}
To introduce our QCPP quantum algorithmic framework, we will start from two opposing directions: first from a top down direction formulating the ultimate computational goal and second from the bottom up direction explaining the basic stochastic building block with the goal of illustrating how it needs to be employed to reach the algorithmic goal. The entire workflow is illustrated in Fig.~\ref{fig:workflow_qcpp}.

\subsection{Algorithmic Goal}
	A huge class of quantum algorithms can be described by applying a function $f$ of a Hermitian operator $H$ on a generic initial state $\rho$ of an $n$-qubit computational register, 
    \begin{equation}
       f(H) \rho f(H)^\dagger = f(\mathcal{A}_H - i \mathcal{C}_H)f^*(\mathcal{A}_H + i \mathcal{C}_H)\rho \,,
    \end{equation}
    followed by a measurement on the resulting state. In anticipation of the introduction of the stochastic building block in the next subsection we reformulate the high level quantum algorithm in the superoperator picture with the help of the commutator $\mathcal{C}_H \rho = \frac{i}{2}[H,\rho]$ and anti-commutator $\mathcal{A}_H \rho = \frac{1}{2}\{H,\rho\}$ superoperators. These commute $\mathcal{C}_H\mathcal{A}_H \rho = \mathcal{A}_H\mathcal{C}_H \rho$ as long as we consider them with respect to the same operator $H$. We consider the Hermitian operator $H$ without loss of generality to be expressed as some real prefactor $\lambda$ and a convex combination of a subset $S \subset G_n$ of signed n-qubit Pauli operators $G_n = \{\pm P_1 \otimes \dots \otimes P_n| P_i \in \{\mathbbm{1}, Z, X, Y\}\}$ where $Z,X$ and $Y$ are the Pauli matrices, $H = \lambda \sum\limits_{g \in S} p_g g$. All $p_g$ are real and positive and sum up to $1$, $\sum_g p_g = 1$. Be aware of the unusual order of Pauli operators, starting with $Z$ and then $X$ and $Y$, that we have deliberately chosen for the paper as it considerably simplifies the derivations to come. 
    
	In the following, we will show how to implement a probabilistic quantum algorithm for functions $f$ that are polynomials of finite degree $d$, $f(z) \propto p(z) =  a_d \prod\limits_{i=1}^d(z-z_i)$, represented here in their factorized form with leading coefficient $a_d$ and roots $\{z_i \in \mathbb{C} | p(z_i) = 0\}$, which is the step 1 in the workflow in Fig.~\ref{fig:workflow_qcpp}. Therefore, for a general function $f$, we need to find an interpolating polynomial $p$ that approximates $f$ on an appropriately defined interval such that,
    \begin{widetext}
    \begin{equation}\label{eq:algo_channel}
        f(H) \rho f(H)^\dagger \approx |a_d|^2 \prod\limits_{i=1}^d (\mathcal{A}_H - i\mathcal{C}_H - z_i)(\mathcal{A}_H + i\mathcal{C}_H - z_i^*)\rho = |a_d|^2 \prod\limits_{i=1}^d \left[(\mathcal{A}_H - \Re[z_i])^2+(\mathcal{C}_H + \Im[z_i])^2\right]\rho =: \mathcal{F}\rho.
    \end{equation}
    \end{widetext}
    In order not to make the nomenclature too cumbersome for the remainder of the article we will consider ``normalized'' Hamiltonians $H\rightarrow h / \lambda$ and assume interpolating polynomials approximating the appropriately rescaled function, $p(\cdot) \approx f(\lambda \cdot)$. With this convention it is sufficient that the interpolating polynomial is approximating the function $f$ well on the  interval $[-1,1]$. 
    
\subsection{Probabilistic Building Block}
    To accomplish this task, we consider a probabilistic mixture of unitary channels, 
    \begin{align}\label{eq: channel_oracle}
        \mathcal{E} = p_z [U_z] + (1-p_z)\sum\limits_{g \in S} p_g [cR_g(\theta)]\,.
    \end{align}
    Here, we are using the convention of square brackets to symbolize the mapping of a unitary operator to its corresponding unitary channel $[U]\rho \coloneqq U\rho U^\dagger$. To implement the desired parametrized unitaries, we supplement a computational register hosting $n$ qubits with an ancillary qubit. With probability $p_z$ we execute a $-\pi/4$ rotation along the $z$-axis on the ancilla qubit $\exp(i\frac{\pi}{4} Z_a)$ and with probability $(1-p_z) p_g$ we execute a rotation with generator $g$ with an angle $\theta$ controlled by the ancilla qubit $cR_g(\theta) = \exp(-i\theta (|0\rangle\langle 0|_a \otimes g) = \exp(-i\theta (\frac{1}{2}(Z_a + 1) \otimes g) $, where we use the subscript $a$ for Pauli operators to symbolize a Pauli operator acting on the ancilla qubit. The circuit diagram of this computational building block is shown in step 2 in Fig.~\ref{fig:workflow_qcpp}.
    
    To better understand the effect of the above defined computational building block on the computational register, we define a partial transfer matrix picture with respect to the ancilla qubit for an arbitrary channel $\mathcal{E}$, $[\mathcal{E}]^{(\textsc{TM})}_{\Sigma,\Delta}\rho = \frac12\text{tr}_a[\Sigma \otimes \mathbbm{1}\, \mathcal{E}(\Delta \otimes \rho)]$ for $\{\Sigma,\Delta\} \in \{\mathbbm{1},Z,X,Y\}^{\otimes 2}$, where $\text{tr}_a[\cdot]$ is the partial trace with respect to the ancilla qubit. With this convention the above introduced algorithmic building block can be written as a direct sum $[\mathcal{E}]^{(\textsc{TM})} =  \mathbf{A}_{\mathbbm{1}-Z}\oplus \mathbf{B}_{X-Y}$ where,
    \begin{widetext}
    \begin{equation}
        \mathbf{A}_{\mathbbm{1}-Z} = 
        \begin{pmatrix}
            \frac{1-p_z}{2}\sum\limits_{g \in S} p_g [R_g(\theta)] + \frac{1 + p_z}{2} &
            \frac{1- p_z}{2}(\sum\limits_{g \in S} p_g [R_g(\theta)] - 1) \\
            \frac{1- p_z}{2}(\sum\limits_{g \in S} p_g [R_g(\theta)] - 1) &
            \frac{1-p_z}{2}\sum\limits_{g \in S} p_g [R_g(\theta)] + \frac{1 + p_z}{2} 
        \end{pmatrix} \,,
    \end{equation}
    acts on the $\mathbbm{1}$-$Z$ subspace and 
    \begin{equation}\label{eq:oracle_matrix}
        \mathbf{B}_{X-Y} = (p_z - 1) \sin(\theta)
        \begin{pmatrix}
            \mathcal{C}_H - \cot(\theta) &
            \mathcal{A}_H + \frac{p_z }{(p_z -1) \sin(\theta)}\\
           -(\mathcal{A}_H + \frac{p_z }{(p_z -1) \sin(\theta)})  &
            \mathcal{C}_H - \cot(\theta) 
        \end{pmatrix} \,,
    \end{equation}
    \end{widetext}
    acts on the $X$-$Y$ subspace, where $R_g(\theta)= \exp(-i\theta g)$ is the uncontrolled version of the unitary of the computational building block.
    
\subsection{Assembling the Building Blocks}
    Next, we show how to assemble the above introduced computational building block such that we are able to implement the desired channel of  Eq.~(\ref{eq:algo_channel}). As a first step we consider for the moment a polynomial of degree one ($d=1$), and recognize that the determinant of $\mathbf{B}_{X-Y}$ is proportional to the desired channel when we choose the angle $\theta \leftarrow \theta_1$ and probability $p_z \leftarrow p_{z_1}$ to be, 
    \begin{align}\label{eq: angle}
        \theta_1 = \begin{cases}
\displaystyle \arctan(\Im z_1) + \frac{\pi}{2},
    & \Re z_1 \geq 0, \\[6pt]
\displaystyle \arctan(\Im z_1) - \frac{\pi}{2},
    & \Re z_1 < 0.
\end{cases}\,,
    \end{align}
     and 
    \begin{align}\label{eq: probability}
        p_{z_1} = \frac{|\Re[z_1]|}{|\Re[z_1]| + \sqrt{1 + \Im[z_1]^2}}\,,
    \end{align}
    which ultimately means that the basic computational building block is now a function of a specific root of the interpolating polynomial $\mathcal{E} \rightarrow \mathcal{E}(z_i)$.
    As a second step, we realize that because of the direct sum structure of the basic computational building block in the partial Pauli transfer matrix picture the same structure also emerges when we start concatenating multiple computational building blocks, 
    \begin{align}
        [\mathcal{E}(z_p) \circ \dots \circ \mathcal{E}(z_1)&]^{(\textsc{TM})} =  \nonumber \\ &\mathbf{A}(z_p)\dots\mathbf{A}(z_1)\oplus \mathbf{B}(z_p)\dots\mathbf{B}(z_1)\,.
    \end{align}
    Combining the above two insights we can conclude that if we could somehow compute the determinant of the $X$-$Y$ subspace of a channel that has the above described direct sum structure we would be able to implement a channel on the computational register proportional to  Eq.~(\ref{eq:algo_channel}) because the determinant of the product of matrices is the product of their determinants, $\det(\mathbf{M}_1\mathbf{M}_2\dots\mathbf{M}_p)= \det(\mathbf{M}_1)\det(\mathbf{M}_2)\dots\det(\mathbf{M}_p)$. This product of the matrices is illustrated in step 3 in Fig.~\ref{fig:workflow_qcpp}.

    To see how we can compute the desired determinants we first define the following product: $D_K = Z \prod\limits_{i \in K }\mathbf{B}_{X-Y}(\theta_i) Z \prod\limits_{i \in K}\mathbf{B}_{X-Y}(\theta_i)$, such that $K$ is a subset of the computational building blocks needed to implement the polynomial of Eq.~(\ref{eq:algo_channel}). Note that for any subset $K$ we have: 
    \begin{align}
        D_K \propto \prod\limits_{i\in K}\left((\mathcal{A}_H - \Re[z_i])^2 + (\mathcal{C}_H + \Im[z_i])^2\right)\mathbbm{1}\,
    \end{align}
    such that the product of all $D_{K_l}$ of any partition $\{K_l\}$ of the original set of $d$ roots of the polynomial would lead to a proper computation of the determinant in the $X$-$Y$ subspace. The Pauli $Z$ gate in the formula for $D_K$ however is on the level of the partial Pauli transfer matrix picture and not on the level of the quantum circuit. To implement a Pauli $Z$ term in the $X$-$Y$ subspace of the partial Pauli transfer matrix picture it suffices on the quantum circuit level to apply an $X$ gate to the ancilla qubit since, $[X_a]^{(\textsc{TM})} = Z_{\mathbbm{1}-Z}\oplus Z_{X-Y}$. With this we can see that we implemented in the $X$-$Y$ subspace a channel that is proportional to the desired application of a polynomial of $H$ on a starting state $\rho$ according to equation~(\ref{eq:algo_channel}).

    As a last step we need to make sure that we can apply the channel in the $X$-$Y$ subspace on the computational register. To this end, we initialize the ancilla qubit in the $|+\rangle$ state and measure in the $X$ basis. Initializing the ancilla qubit in the $|+\rangle$ state is in the partial Pauli transfer matrix picture equivalent to applying all matrices on the initial state, $[|+\rangle\langle + |_a]^{(\textsc{TM})}= [\frac{1}{2}(\mathbbm{1} + X_a)]^{(\textsc{TM})} = (\frac{1}{2}, 0, \frac{1}{2}, 0)^T$ while measuring in the $X$-basis of the ancilla qubit is equivalent to collapsing the state into either $[\langle+ | \cdot | +\rangle]^{(\textsc{TM})} = (\frac{1}{2}, 0 ,\frac{1}{2}, 0)$ or $[\langle - | \cdot | - \rangle]^{(\textsc{TM})} = (\frac{1}{2}, 0 ,-\frac{1}{2}, 0)$. The measurement probabilities of the ancilla qubit in the $X$ basis generally depend on the input state and the applied quantum channel. The two measurement outcomes correspond to unnormalized conditional states on the computational register, which can be expressed in the partial Pauli transfer matrix representation as follows:
    \begin{widetext}
    \begin{align}\label{eq: final_measurement}
    &\rho \rightarrow 2 \begin{pmatrix}
            \frac12 & 0 & \pm\frac12 & 0 
        \end{pmatrix} \begin{pmatrix} Z \mathbf{A}(\theta_d) \dots \mathbf{A}(\theta_1)Z \mathbf{A}(\theta_d) \dots \mathbf{A}(\theta_1)  & 0_{2\times2} \\0_{2\times2}  & Z \mathbf{B}(\theta_d) \dots \mathbf{B}(\theta_1)Z \mathbf{B}(\theta_d) \dots \mathbf{B}(\theta_1)  \end{pmatrix}  \begin{pmatrix}
            \frac12 \\ 0 \\ \frac12 \\ 0 
        \end{pmatrix} \rho \\
       & \rho \rightarrow \frac{1}{2}\left(\underbrace{\begin{pmatrix}
            1 & 0 
        \end{pmatrix} Z \mathbf{A}(\theta_d) \dots \mathbf{A}(\theta_1)Z \mathbf{A}(\theta_d) \dots \mathbf{A}(\theta_1) 
        \begin{pmatrix}
            1 \\0
        \end{pmatrix}}_{\mathcal{P}} \pm 
        \underbrace{
        \begin{pmatrix}
            1 & 0 
        \end{pmatrix} Z \mathbf{B}(\theta_d) \dots \mathbf{B}(\theta_1)Z \mathbf{B}(\theta_d) \dots \mathbf{B}(\theta_1) 
        \begin{pmatrix}
            1 \\0
        \end{pmatrix}}_{\propto \mathcal{F}}  \right)\rho.
    \end{align}\,
    \end{widetext}
    This step of the QCPP workflow is illustrated in step 4 in Fig.~\ref{fig:workflow_qcpp}. However, the above formula can be generalized to any other partition $\{K_l\}$. This means that if we were to ignore the measurement results of the ancilla qubit, the desired channel $\mathcal{F}$ would average out. In an unraveling of the actual quantum measurement of an observable on the computational register $O$ we would measure its eigenstates $|o\rangle$ which do not correspond to the probabilities we would want to have, namely $p_o \propto \text{tr}[| o\rangle\langle o| \mathcal{F}\rho]$, but rather those of $q_o \propto [|o\rangle\langle o|\mathcal{P}] \pm p_o$, depending on this measurement outcome of the ancilla. It is precisely the measurement outcome of the ancilla qubit that lets us tag the positive and negative outcomes which we can subsequently use to change the sign of the eigenvalue corresponding to the measured eigenstate of the observable to ultimately average out the unwanted channel $\mathcal{P}$. 
\section{Sample- and Query Complexity Tradeoff}\label{sec:sample_query} 
To estimate the mean value of an observable with the above introduced stochastic framework we first need to calculate the actual factor of proportionality between the channel we want to implement and the channel we can implement which ultimately is the product of all prefactors of the oracle calls of equation~ (\ref{eq:oracle_matrix}), 
\begin{equation}\label{eq: sampling_complexity}
\Gamma(p) := |a_d| \prod\limits_{i = 1}^d (|\Re[z_i] |+ \sqrt{1 + (\Im[z_i])^2}).
\end{equation}
$\Gamma(p)$ determines the statistical overhead of the stochastic reconstruction. For an observable $O$ with $|O|\leq 1$, assigning a sign $(s=\pm1)$ according to the ancilla measurement outcome yields a signed estimator with expectation value $\operatorname{tr}[O\mathcal{F}(\rho)]/\Gamma(p)^2$. Consequently, an additive statistical error $\epsilon_{\mathrm{stat}}$ in the target expectation can be achieved using $M=O(\Gamma(p)^4/\epsilon^2)$ independent circuit samples at constant success probability. In the following, we refer to $\Gamma(p)$ as the sample complexity, which characterizes the statistical cost of QCPP. In addition, the query complexity is determined by the number (d) of basic computational building blocks, which directly relates to the quantum circuit depth. In the following, we argue that the optimal query complexity can be achieved for QCPP, but only at the cost of an exponentially growing sample complexity. We go on to construct a class of approximating polynomials that enable a variable tradeoff between sample and query complexity. Ultimately, we can show that as a function of query complexity we are able to get a exponentially decreasing additive error as well as an only polynomially growing sample complexity. Specifically, we generate these two insights for both real and imaginary time evolution. 

\subsection{Optimal Query Complexity for Real and imaginary Time Evolution}
It is possible to get the optimal query complexity for real-, $f^{\text{real}}(z) = \exp(- i t z)\approx p^{\text{real}}(z)$, and imaginary, $f^{\text{imag}}(z) = \exp(- \tau (z + 1)) \approx p^{\text{imag}}(z)$, time evolution in terms of the respective simulated time, $t$ or $\tau$, as well as in terms of the additive error of the approximation. This can be accomplished by using, for example, the respective Jacobi-Anger expansions, 
\begin{align}\label{eq:JA_expansion}
p_{\textrm JA}^{\mathrm imag}(z) &=  e^{-\tau}(I_0(\tau)+2\sum_{n=1}^d(-1)^nI_n(\tau)T_n(z))\\
p_{\textrm JA}^{\mathrm real}(z) &=J_0(t)+2\sum_{n=1}^d(-i)^nJ_n(t)T_n(z)\,,
\end{align}
where $J_n$ are the ordinary Bessel functions of first kind and $I_n$ are the modified Bessel functions of first kind and $T_n$ are the Chebyshev polynomials of first kind. However, as mentioned above we were able to gather numerical evidence that the sample complexity grows exponentially in $d$. Specifically we are able to show an exponentially growing upper bound. 

\begin{theorem}\label{thm:exp_growth}
\longthmtitle{Exponential sample complexity of Jacobi-Anger expansions}
Let $p^{\mathrm real}(x)$($p^{\mathrm imag}(x)$) be the Jacobi-Anger expansion of real(imaginary) time evolution truncated at degree $d$. Then it is possible to find constants $c^{\textrm real}$ and $c^{\textrm imag}$, independent of $d$ such that the sample complexity is growing at most exponentially in $d$, 
\begin{align}
\Gamma(p^{\textrm real}) \in O(\exp(c^{\textrm real} d))\\
\Gamma(p^{\textrm imag}) \in O(\exp(c^{\textrm imag} d))\,,
\end{align}
for real- and imaginary time evolution respectively.
\end{theorem}
For details of the proof we refer to the supplementary material. We further provide numerical evidence that the exponentially growing upper bound is tight, c.f. appendix section~\ref{{app_sec:lower_bound_numerics}}. An important insight here is that the sample complexity is growing  exponentially in the number of roots of the interpolating polynomial that are neither lying on the positive real axis nor on the imaginary axis of the complex plane. We use this insight to construct interpolating polynomials in the next section that provide us with a flexible tradeoff between sample- and query complexity.

\subsection{Flexible Tradeoff between Sample- and Query Complexity}
We construct interpolating polynomials as products of two polynomials $p(z) = q(z) h(z)$. For this product we see that the sum of the degrees of the polynomials $\deg(q) = d_q$ and $\deg(h) = d_h$ is $d$ and the sample complexity is the product of the sample complexities of the two polynomials, $\Gamma(p) = \Gamma(q)\Gamma(h)$. For $q$ we choose a class of polynomials whose sample complexity is equal to $1$. For the case of real- and imaginary time evolution these are, respectively, 
\begin{align}
 q^{\textrm real}(z)&=\frac{(1-i \frac{t}{d_q} z)^{d_q}}{\left(1+\frac{t^2}{d_q^2}\right)^{d_q/2}}\label{eq: real_protected_factor}\\
    q^{\textrm imag}(z)&= \frac{(1-\frac{\tau}{d_q}z)^{d_q}}{(1+\frac{\tau}{d_q})^{d_q}} \label{eq: imag_protected_factor}\,.
\end{align}
In the case of real time evolution they are all the same and located on the imaginary axis $\{z_i\}_{i=1}^{d_q} = -i d_q/t$. For imaginary time they are all identical as well and located on the positive real axis $\{z_i\}_{i=1}^{d_q} = d_q/\tau$. The remaining polynomial $h$ we define to be the approximation of $h^{\textrm imag} \approx f^{\textrm imag} / q^{\textrm imag}$ and $h^{\textrm real} \approx f^{\textrm real} / q^{\textrm real}$ based on Chebyshev polynomials truncated at degree logarithmic in $d$, $d_h \propto \log(d)$, respectively. With this construction it is obvious that the sample complexity is polynomial in the query complexity, that means the degree of the interpolating polynomial $d$. We were further able to show exponential convergence for these specific polynomials on the relevant interval of $[-1,1]$. 

\begin{theorem}\label{thm: query-sample}
    \longthmtitle{Query-sample complexity tradeoff}
    Let $q^{\mathrm{real}(\mathrm{imag})}$,
    $h^{\mathrm{real}(\mathrm{imag})}$, and
    $f^{\mathrm{real}(\mathrm{imag})}$ be defined
    as above. Within the asymptotic regime
    specified in Appendix~\ref{section: appendix sample-query tradoff}, the polynomial
    degree $d$, corresponding to the query
    complexity, scales as
        \begin{align}
        d &= \mathcal{O}\left(\frac{t^2}{[\log\Gamma_{\star}]^3}\right)+\mathcal{O}\left(\log\frac{1}{\epsilon}\right),
        \\
        d &= \mathcal{O}\left(\frac{\tau^2}{[\log \Gamma_{\star}]^3}\right)+\mathcal{O}\left(\log\frac{1}{\epsilon}\right),
\end{align}
for real and imaginary time evolution. Here, $\Gamma_\star$ is the upper bound of the sample complexity and the upper bound of the approximation error $\epsilon$ is defined as
\begin{align}
    \epsilon:&=\|f^{\textrm real}(t\cdot)- q^{\textrm real}h^{\textrm real} \|_{\infty}\\
    \epsilon:&=\|f^{\textrm imag}(\tau\cdot)- q^{\textrm imag}h^{\textrm imag} \|_{\infty}
\end{align}
where $\|f(x)\|_{\infty} = \sup_{x\in[-1,1]}|f(x)|$ is the supremum norm on the interval $[-1, 1]$. 
\end{theorem}

For the proof of the theorem we refer to the supplementary material section~\ref{section: appendix sample-query tradoff} and the upper bound of the query complexity is supported by our numerical data as detailed in the supplementary material \ref{app:numerics}.

\section{General Remarks} \label{sec:general_remarks}
\subsection{Commutative Structure}
Note that basic computational building blocks do not commute for general roots $\mathcal{E}(z_1)\mathcal{E}(z_2) \neq \mathcal{E}(z_2)\mathcal{E}(z_1)$. However, a permutation of the computational building blocks order does not have any effect on the relevant part of the quantum channel acting on the computational register. That means we are still implementing the desired channel $\mathcal{F}$ on the computational register. This can be seen from at least two possible angles. First, it can be seen from the fact that the desired channel $\mathcal{F}$ is a result of taking the determinant of a set of computational buildings blocks and the fact that the determinant is actually permutation invariant for matrix products, $\det(\mathbf{M}_1\mathbf{M}_2) = \det(\mathbf{M}_2\mathbf{M}_1)$. Second, we can also see that every application of the basic computational building block is associated to a root of the polynomial we want to apply and the roots have no inherent order. This fact can be used to find more favorable transpilation techniques for various quantum hardware platforms. However, it might even hint at more fundamental insights about the general QCPP algorithm, similar to findings about instantaneous quantum polynomial circuits \cite{shepherdTemporallyUnstructuredQuantum2009,bremnerClassicalSimulationCommuting2010,bremnerAverageCaseComplexityApproximate2016}. 
In contrast to the desired channel $\mathcal{F}$, the other channel $\mathcal{P}$ actually is not independent of the chosen sequence of the basic computational building blocks. The exact channel can be conveniently calculated by first considering the action of the entire quantum algorithm on an initial state that is a tensor product of the projectors onto computational basis states of the ancilla qubit, $|0/1\rangle\langle0/1|\otimes\rho$. For every partition we realize that, thanks to the controlled rotation $\mathcal{E}(z_i)$, only one of the two copies of the basis computational building blocks is actually acting on the computational register. Ultimately, we can determine the channel $\mathcal{P}$ to be, 
\begin{equation}
\mathcal{P} = \mathcal{E}'(z_d) \circ \dots \circ\mathcal{E}'(z_1)\,,
\end{equation}
where we have introduced,
    \begin{align}
        \mathcal{E}' = p_z\mathbbm{1}+ (1-p_z)\sum\limits_{g \in S} p_g [R_g(\theta)]\,,
    \end{align}
as the uncontrolled version of the basic computational building block. 
\subsection{Comparison with Quantum Singular Value Transformations}
The here presented quantum algorithmic framework bears close resemblance with the quantum algorithmic framework of Quantum Singular Value Transformations (QSVT). With QSVT one also has the ability to apply polynomials of linear operators onto an initial state. In contrast, however, to the here presented stochastic framework, QSVT is based on the idea of quantum signal processing (QSP) and general quantum embeddings. The basic insight of the QSP protocol is that one can implement almost arbitrary polynomials $p(H)$ of a Hermitian operator $H$ with a concatenation of a signal rotation operator $W(H)$ and a quantum signal processing rotation $e^{i\phi Z}$, 
\begin{equation}
P(H) = \langle 0 | e^{i \phi_0 Z} \prod\limits_{k=1}^d W(H) e^{i\phi_k Z} | 0 \rangle\,. 
\end{equation}
 In the context of this work the Hermitian operator could be a Hamiltonian governing, for example, the real time evolution of a system of interest, while the polynomial would be approximating the complex exponential function $P(z)\approx \exp(-itz)$. The approximate real time evolution would then proceed by applying the quantum signal processing sequence on an arbitrary initial state $|\psi_0\rangle$ and herald on the $0$ measurement,
\begin{equation}
|\psi_t\rangle \approx \langle 0| e^{i \phi_0 Z} \prod\limits_{k=1}^d W(H) e^{i\phi_k Z}    |0\rangle \otimes |\psi_0\rangle\,.
\end{equation}
Instrumental to this technique is our ability to implement  a unitary operator $W(H)$ such that $H = \langle 0 | W(H)| 0 \rangle$, which is the topic of block encodings. One possibility to accomplish a block encoding for a Hamiltonian which can be described as a sum of Pauli operators is the technique of linear combination of unitaries (LCU). We need an ancillary qubit register of logarithmic size in the number of terms of the Hamiltonian and to prepare a state $|\Phi\rangle = \sum\limits_{g\in S}\sqrt{p_g} |g\rangle$. We complement the ancillary qubit register with the computational qubit register of size $n$ and apply the unitary Pauli operators of the Hamiltonian, controlled on the ancillary qubit register, 
\begin{equation}
|\Phi\rangle\otimes|0\rangle\rightarrow \sum\limits_{g\in S} \sqrt{p_g} |g\rangle\otimes g|0\rangle.
\end{equation}
We finally apply the inverse of the unitary that prepared the state $|\Phi\rangle$ on the ancillary register. The entire procedure implements the correct signal rotation operator. In contrast to QCPP we need an ancillary qubit register of size logarithmic in the number of terms of the Hamiltonian rather than simply a single ancilla qubit. This is still efficient in a complexity theoretic sense and yet it is daunting to implement on a NISQ quantum computer. 

Comparing QSVT to the here presented stochastic framework, we can conclude that the ``signal'', that means the information of the actual Hermitian operator we want to apply, is encoded in the probabilities with which we apply the controlled rotations generated by Pauli operators  rather than the state of an ancillary qubit register. In turn the information about which polynomial of the Hermitian operator we want to apply is encoded, first in the angles of the controlled rotations and second as the probabilities with which we apply the $Z$ rotations on the single ancilla qubit that is needed for the stochastic framework, $\exp(i\frac{\pi}{4} Z_a)$. 

\section{Conclusion}\label{sec:conclusion}

We have developed QCPP, a circuit-sampling approach for implementing a quantum channel that applies a polynomial of a Hamiltonian to an initial state. Each basic computational building block requires only controlled Pauli-rotation circuits for the Pauli terms appearing in the problem Hamiltonian. We have also identified the exponential sample complexity associated with the Jacobi-Anger expansion for target functions such as real-time and imaginary-time exponential functions. To overcome this obstruction, we constructed polynomials that realize a sample-query complexity trade-off: the sample complexity remains polynomially bounded while the query complexity becomes quadratic in evolved time instead of the optimal query complexity, that is linear in evolved time, meanwhile the exponential convergence of the approximation error in terms of the query complexity is retained. These results provide a route toward quantum channel polynomial processing on NISQ and early fault tolerant quantum computers. Future work will demonstrate the experimental implementation of the algorithm on quantum hardware and extend the theoretical framework to interpolating a broader class of target functions using polynomially bounded circuit sampling cost.

\section*{Acknowledgements}
We thank St\'ephanie Cheylan, Oriel Kiss and  Jinzhao Sun for useful and insightful discussions.

\bibliography{ref}

\newpage
\onecolumngrid
\appendix

\section{Upper Bound of the Sample Complexity}\label{{app_sec:lower_bound_numerics}}
Here we provide  in corollary~\ref{corr:Chebyshevl1} an upper bound on the sample complexity for implementing an arbitrary polynomial and with that we provide the proof for the exponential scaling of the upper bound on the sample complexity for the interpolating polynomials for real and imaginary time evolution that we get from a truncated Jacobi-Anger expansion. 

\begin{theorem}\label{theorem:upper_bound}
    Assume the truncated Jacobi-Anger expansion at degree $d$ for real(imaginary) time evolution $p_{\textrm JA}^{\textrm real(imag)}$ introduced in the main text Eq.~\ref{eq:JA_expansion}. 
   Then the sample complexity has an exponentially scaling upper bound,
   \begin{align}
    \Gamma(p_{\textrm JA}^{\textrm real})&\leq (2+2\sqrt{2})^d(1+\sqrt{2d})\\
    (\Gamma(p_{\textrm JA}^{\textrm imag})&\leq (2+2\sqrt{2})^d)
   \end{align}
\end{theorem}
To prepare for the proof we begin by introducing some necessary nomenclature in the following definition.
\begin{definition}[Power and Chebyshev basis $\ell^1$ norms]
Let $p$ be a polynomial of degree $d$. Suppose it is written in the power basis as $p(x)=\sum_{\ell=0}^{d} a_\ell x^\ell$ and in the Chebyshev basis as $p(x)=\sum_{m=0}^{d} t_m T_m(x)$ where $t_m$ is the $m$-th Chebyshev coefficient and $T_m$ is the $m$-th Chebyshev polynomial of the first kind. Define the $\ell^1$ norm in the power and Chebyshev basis as:
\begin{align}
  \|p\|_{ \mathrm{pow}}:=\sum_{\ell=0}^{d}|a_\ell|,
  \qquad
  \|p\|_{T}:=\sum_{m=0}^{d}|t_m|.    
\end{align}
The phrase ``degree $d$'' below means degree at most $d$; if the true degree is smaller, the unused top coefficients are zero.
\end{definition}
The proof of theorem \ref{theorem:upper_bound} builds on a first lemma~\ref{lemma: sample_cost_pwer_norm} that introduces an upper bound of the sample complexity proportional to the $\ell^1$-norm in power basis. We further set up a second lemma~\ref{lemma:Chebyshev_to_power} that establishes an upper bound for the $\ell^1$-norm in the power basis proportional to the $\ell^1$-norm in the Chebyshev basis. To finally prove the theorem, what remains to be done is simply estimating the $\ell^1$ norm in the Chebyshev basis for the truncated Jacobi-Anger expansion of real and imaginary time evolution.  
\begin{lemma}[Upper bound on sample cost proportional to \(\ell^1\)-norm in the power basis]\label{lemma: sample_cost_pwer_norm}
Let $p(z)=\sum_{\ell=0}^{d}a_\ell z^\ell=a_d\prod_{j=1}^{d}(z-z_j)$ be a nonzero complex polynomial of degree \(d\). Then
\begin{align}
    \boxed{
    \Gamma(p)\le 2^d\|p\|_{\mathrm{pow}}
    }
\end{align}
\end{lemma}
\begin{proof}
The sample complexity can be upper bounded with the help of the coefficient norm in the power basis of a polynomial. To this end we first prove the following upper bound for the factors of the sample complexity
\begin{equation}
    |x|+\sqrt{1+y^2} \le 2\max\{1,|z|\}\,,
\end{equation}
for \(z=x+\mathrm{i}y\). If \(|z|\le1\), then \(x^2+y^2\le1\), so \(|x|\le \sqrt{1-y^2}\). Therefore 
\begin{equation} \label{eq: inequality |z|<1}
    |x| + \sqrt{1 + y^2}\le\sqrt{1-y^2}+\sqrt{1+y^2}\le \sqrt{2[(1-y^2) + (1 + y^2)]} = 2,
\end{equation}
where the last inequality holds because the arithmetic mean is less or equal to the quadratic mean. The equality holds when $y=0$. Thus \(|x|+\sqrt{1+y^2}\le2=2\max\{1,|z|\}\) in this case.

If \(|z|\ge1\),  \(y^2=|z|^2-x^2\), and therefore
\begin{align}\label{eq: inequality |z|>1}
    |x| + \sqrt{1 + y^2}
    =
    |x|+\sqrt{1+|z|^2-x^2}
    \le
    |x|+\sqrt{2|z|^2-x^2}
    \le
    \sqrt{2(x^2 + 2 |z|^2 - x^2)} = 2|z|
\end{align}
In the last inequality we again use the fact that the arithmetic mean is smaller or equal to the quadratic mean. Hence from Eq.~\ref{eq: inequality |z|<1} and Eq.~\ref{eq: inequality |z|>1}, we have shown that $|\Re [z]|+\sqrt{1+(\Im[z])^2}\le 2\max\{1,|z|\}$ for all complex numbers $z$.  
This means  the sample complexity is upper bounded by $\Gamma(p)\le2^d |a_d|\prod_{j=1}^{d}\max\{1,|z_j|\}=2^d \mathbb{M}(p)$ where $\mathbb{M}(p):=|a_d|\prod_{j=1}^{d}\max\{1,|z_j|\}$ is the Mahler measure of the polynomial \(p\) \cite{Mahler1962Inequalities, EverestWard1999Heights}.  It remains to compare \(\mathbb{M}(p)\) with the coefficient \(\ell^1\)-norm~\cite{EverestWard1999Heights}. By Jensen's formula and one inequality on the unit circle,
\begin{eqnarray}
    \begin{split}
    \mathbb{M}(p)=&\exp\left(\frac{1}{2\pi}\int_0^{2\pi}\log|p(e^{\mathrm{i}\theta})|\,d\theta\right).\\
        |p(e^{\mathrm{i}\theta})|
    \le&\sum_{\ell=0}^{d}|a_\ell||e^{\mathrm{i}\ell\theta}|=\sum_{\ell=0}^{d}|a_\ell|=\|p\|_{\mathrm{pow}}.
    \end{split}
\end{eqnarray}
Therefore $\log \mathbb{M}(p)\le\log\|p\|_{\mathrm{pow}},$ and hence $\mathbb{M}(p)\le\|p\|_{\mathrm{pow}}.$
Combining this with \(\Gamma(p)\le2^d\mathbb{M}(p)\) provides us with,
\begin{align}
    \Gamma(p)
    \le
    2^d\|p\|_{\mathrm{pow}}.
\end{align}
This proves the claim.
\end{proof}
We go on proving our second lemma establishing an upper bound for the $\ell^1$-norm in the power basis proportional to the $\ell^1$-norm in the Chebyshev basis. 
\begin{lemma}[Upper bounding power- by Chebyshev $\ell^1$ norm]\label{lemma:Chebyshev_to_power}
Let $p(x)=\sum_{m=0}^{d}t_mT_m(x)=\sum_{\ell=0}^{d}a_\ell x^\ell$ be a polynomial of degree $d$. Then
\[
  \|p\|_{\mathrm{pow}}
  \le (1+\sqrt2)^d\|p\|_{T}.
\]
\end{lemma}
\begin{proof}
One can show that for every $m\ge 0$,
\[
  \|T_m\|_{\mathrm{pow}}\le (1+\sqrt 2)^m.
\]
The first two Chebyshev polynomials are $T_0(x)=1$ and  $T_1(x)=x,$ so $\|T_0\|_{\mathrm{pow}}=1$ and$\|T_1\|_{\mathrm{pow}}=1.$
For $m\ge 1$, the Chebyshev recurrence is  $T_{m+1}(x)=2xT_m(x)-T_{m-1}(x).$
Multiplication by $x$ maps the coefficients in the following way $a_i \rightarrow a_{i+1}$, so it preserves the
power $\ell^1$ norm: $\|xP(x)\|_{\mathrm{pow}}=\|P(x)\|_{ \mathrm{pow}}.$
Therefore, by the triangle inequality,
\begin{align}
  \|T_{m+1}\|_{\mathrm{pow}} \le 2\|xT_m\|_{\mathrm{pow}}+\|T_{m-1}\|_{1, \mathrm{pow}}=2\|T_m\|_{ \mathrm{pow}}+\|T_{m-1}\|_{\mathrm{pow}}.    
\end{align}
The dominant eigenvalue of the characteristic polynomial of the recurrence inequality $A_{m+1} \leq 2A_m+A_{m-1}$ is $\kappa=1+\sqrt 2$ . Since $\|T_0\|_{ \mathrm{pow}}=1=\kappa^0$ and $\|T_1\|_{\mathrm{pow}}=1\leq \kappa^1$, it is straightforward to prove by induction that:
\begin{align}
  \|T_{m+1}\|_{ \mathrm{pow}}
  \le 2\|T_m\|_{ \mathrm{pow}}+\|T_{m-1}\|_{ \mathrm{pow}}
  \le 2\kappa^m+\kappa^{m-1}
  =\kappa^{m-1}(2\kappa+1)
  =\kappa^{m-1}\kappa^2
  =\kappa^{m+1}.    
\end{align}
Thus $\|T_m\|_{ \mathrm{pow}}\le \kappa^m=(1+\sqrt2)^m$ for all $m\ge0$.

Using the Chebyshev expansion and the triangle inequality in the power basis,  $\|p\|_{ \mathrm{pow}}=\left\|\sum_{m=0}^{d}t_mT_m\right\|_{\mathrm{pow}}  \le \sum_{m=0}^d|t_m|\,\|T_m\|_{1, \mathrm{pow}}.$ From above,  $\|T_m\|_{1, \mathrm{pow}}\le (1+\sqrt2)^m.$
Since $m\le d$, we also have  $(1+\sqrt2)^m\le (1+\sqrt2)^d.$ Therefore
\begin{align}
  \|p\|_{\mathrm{pow}}
  \le \sum_{m=0}^{d}|t_m|(1+\sqrt2)^m
  \le (1+\sqrt2)^d\sum_{m=0}^{d}|t_m|=(1+\sqrt2)^d\|p\|_{T}.    
\end{align}
\end{proof}
We get the following corollary as a result of Lemma~\ref{lemma: sample_cost_pwer_norm} and Lemma~\ref{lemma:Chebyshev_to_power}. It relates the sample complexity upper bound of implementing an arbitrary polynomial with its Chebyshev norm, which we will use later to design interpolating polynomials with polynomially bounded sample complexity.
\begin{corollary}[Sample complexity upper bounded by Chebyshev $\ell^1$ norm]\label{corr:Chebyshevl1}
    The sample complexity of implementing the $d-$degree polynomial $p$ is upper bounded by:
    \begin{equation}\label{eq: sample_cost_upper_bound_Chebyshev_basis}
    \boxed{        \Gamma(p)\leq (2+2\sqrt2)^d\|p\|_{T}}
    \end{equation}
    
\end{corollary}


\begin{proof}[Proof of Theorem~\ref{theorem:upper_bound}]
We can calculate the upper bound in Eq.~\ref{eq: sample_cost_upper_bound_Chebyshev_basis} of the Jacobi-Anger expansion in Eq.~\ref{eq:JA_expansion} of the two exponential functions and get:
\begin{eqnarray}
    \begin{split}
        &\Gamma(p^{\textrm real}_{JA})\leq (2+2\sqrt2)^d[|J_0(t)|+2\sum_{m=1}^d |J_m(t)|]\leq(2+2\sqrt2)^d(1+\sqrt{2d})\\
        &\Gamma(p^{\textrm imag}_{JA})\leq (2+2\sqrt2)^de^{-\tau}[|I_0(t)|+2\sum_{m=1}^d |I_m(t)|]\leq(2+2\sqrt2)^d\\
    \end{split}
\end{eqnarray}
We have used in the first inequality the relation 
\[
(J_0(t))^2+2\sum_{n=1}^{\infty}(J_n(t))^2=1 \Rightarrow \frac2d\sum_{m=1}^d |J_m(t)|\leq \sqrt{\frac2d(\sum_{m=1}^d (J_m(t))^2)} \leq \sqrt{\frac{2}{d}}
\]
and used in the second inequality
\[
e^{\tau}=I_0(\tau)+2\sum_{n=1}^{\infty}I_n(\tau)\geq |I_0(\tau)|+2\sum_{m=1}^d |I_m(\tau)|
\]
Therefore, the upper bound of the sample complexity of implementing the $d-$degree Jacobi-Anger expansion polynomial grows exponentially in terms of $d$. 
\end{proof}

\section{Exponentially scaling Sample Complexity for Jacobi-Anger expansion}
\label{app_sec:lower_bound_numerics}
\begin{figure}[htp]
\centering
\includegraphics[width=0.96\linewidth]{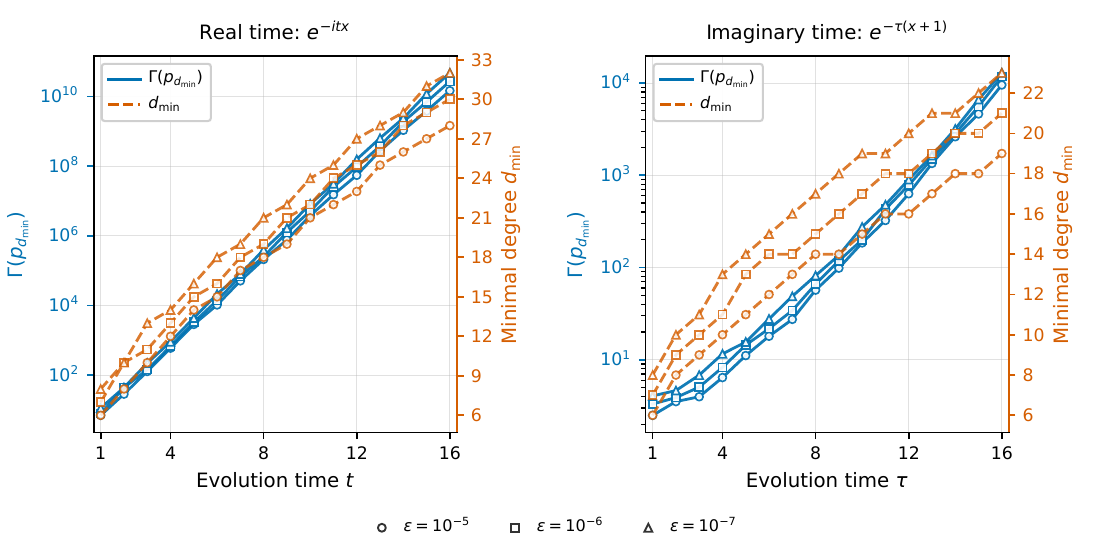}
\caption{The minimal degree of the Jacobi-Anger series and the correspondent sample complexity in terms of evolved time $t$ and $\tau$ to approach the target function of real time evolution $\exp(-itx)$ and imaginary time evolution $\exp(-\tau(x+1))$ with different fixed interpolation residual error $\epsilon$.}
\label{fig: overhead_real_imag_time_evolution}
\end{figure}

We numerically determined the roots of the truncated Jacobi-Anger expansion of real and imaginary time evolution via the Chebyshev colleague matrix with multi-precision method~\cite{Trefethen2019ATAP}. With these numerically calculated roots we get the sample complexity in this section according to Eq.~\ref{eq: sampling_complexity}. 
In Fig.~\ref{fig: overhead_real_imag_time_evolution}, we have plotted, in terms of evolved time $t$ and $\tau$, the minimal degree $d_{min}$ of the Jacobi-Anger series and the corresponding sample complexity $\Gamma(p_{d_{min}})$ to reach interpolation residual errors $\epsilon=10^{-5}, 10^{-6}, 10^{-7}$ respectively.  We see a near linear dependence of the query complexity $d_{min}$ as a function of the evolved time in accordance with the no-fast forwarding theorem~\cite{Berry2007NoFastForwarding}. Meanwhile, the sample complexity $\Gamma(p_{d_{min}})$ grows exponentially in terms of $t$ or $\tau$. The difference in sample complexity for the two exponential functions is due to the asymmetric form in terms of the real and imaginary part of the roots in Eq.~\ref{eq: sampling_complexity}. The difference in minimal query complexity to reach the required residual error is due to the different behavior of Bessel functions as Chebyshev coefficients in the Jacobi-Anger series.

\begin{figure}[htp]
\centering
\includegraphics[width=0.96\linewidth]{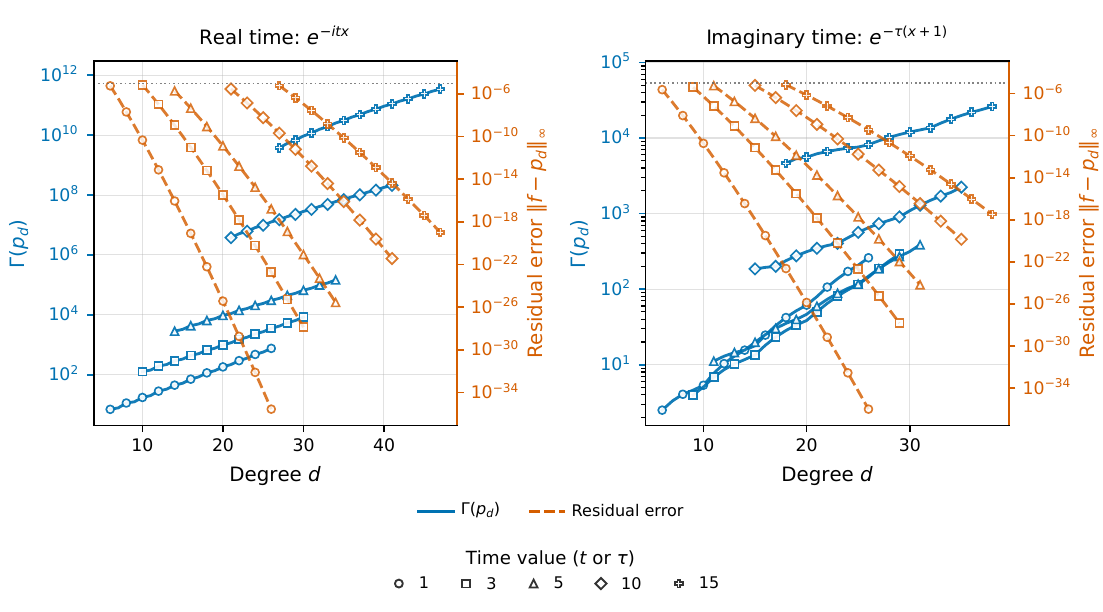}
\caption{The sample complexity and the residual error in terms of the degree of the Jacobi-Anger series for the real and imaginary time evolution functions. We start plotting both quantities only after when the Jacobi-Anger series interpolates the two target functions below a residual error of $\epsilon=10^{-5}$.}
\label{fig: overhead_residual_error_query}
\end{figure}

We have plotted as well in Fig.~\ref{fig: overhead_residual_error_query} the sample complexity $\Gamma(p_d)$ and residual error $\epsilon$ in terms of the query complexity $d$ at different evolved time. Curves are only plotted after the Jacobi-Anger series reach a approximation error of $\epsilon=10^{-5}$ or smaller. We observe an exponential decay of the approximation error and the exponential growth of the sample complexity in terms of query complexity $d$. This numerically indicates that the exponential growth of sample complexity in terms of query complexity is inherent. As long as the query complexity has to grow linearly with evolved time, the sample complexity has to grow exponentially in terms of evolved time.

Conjecturing, for the moment, the sample complexity to grow exponentially we can show that a considerable part of the roots of the truncated Jacobi-Anger series have to reside neither on the positive real or imaginary axis, $E = i \mathbb{R} \cup \mathbb{R}_+$. This insight will ultimately guide us in the next section to construct interpolating polynomials with polynomially scaling sample complexity. Since the truncated Jacobi-Anger series interpolate real and imaginary time evolution we can conclude that $p^{\text{imag/real}}(-1) = \text{const}$. This means we can further state that the following quantity is exponentially growing, 
\begin{equation}\label{eq: root_dist_ratio}
|\frac{\Gamma(p)}{p(-1)}|=\prod_{i=1}^d \frac{|\Re [ z_i]|+\sqrt{1+(\Im [z_i])^2}}{|1+z_i|}\,.
\end{equation}
Every factor in the product is equal to $1$ if the respective root $z_i$ is in $E$, $z_i\in E$. Which leads us to conclude that there is at least polynomial fraction of roots not fulfilling this criterion for $p^{\text{imag/real}}$. In the following we will construct interpolating polynomials with only logarithmically small number of roots that are not in $E$. This argument can be extended if one can find for the argument of $p(\cdot)$ other points than $-1$ such that we can show that the factors in the above ratio, Eq.~\ref{eq: root_dist_ratio} also are equal to 1. 

\section{Flexible Tradeoff between Sample- and Query Complexity}\label{section: appendix sample-query tradoff}
Based on the insights of the previous section, specifically the  upper bound of the sample complexity in terms of the Chebyshev $1-$norm in corollary~\ref{corr:Chebyshevl1}, we develop in this section a class of interpolating polynomials with super-algebraically small approximation error as well as a polynomially growing sample complexity. To make this appendix self contained we will first recall basic insights of the approximation errors of Chebyshev polynomials. We move on to prove super-exponential convergence and exponential sample complexity with Chebyshev interpolants for target functions with mild additional constraints on their growth in the complex plane. This proof indicates with logarithmic $d$ a super-algebraic convergence with polynomial sample complexity. Then, we highlight the role of a ubiquitous scaling parameter in any quantum algorithmic framework and discuss in what sense this provides a challenge for a practical tradeoff between query and sample complexity in the case of QCPP. Specifically, the complication arises out of the requirement of polynomially growing query complexity as a function of a scaling parameter that in the case of real and imaginary time evolution is proportional to the time. This motivates our  specific construction of the interpolation polynomials as a product of polynomials $p = qh$. $q$ serves here as a, sample complexity neutral $\Gamma(q) = 1$, preconditioning polynomial that removes to a certain extent the dependence of sample- and query complexity on the scaling parameter. $h$ is a Chebyshev refinement interpolant, which helps us attain super-algebraic convergence in general and exponential convergence for the real and imaginary time evolution. Our line of argumentation culminates in a specific construction method for $q$ that fulfills all the above introduced requirements for real and imaginary time evolution. 

\subsection{The Bernstein Ellipse and Convergence of Chebyshev Interpolants}\label{appendix section: Bernstein ellipse}

The Bernstein ellipse is an ellipse in the complex plane with foci at \(\pm1\), that is parametrized by $\mathcal{E}_\rho:=\left\{\frac12\left(w+w^{-1}\right): |w|=\rho\right\}$ for some \(\rho>1\). As it turns out 
the convergence of a Chebyshev interpolant towards an arbitrary target function $g$ on the interval $[-1,1]$ is governed by the behaviour of $g$ in the complex plane on the Bernstein ellipse according to the following theorem

\begin{theorem}
    [Bernstein-ellipse bounds for Chebyshev interpolants
{\cite[Theorems~8.1 and~8.2]{Trefethen2019ATAP}}]\label{theorem: Bernstein_ellipse}
Suppose that \(g(z)\) is analytic inside a Bernstein ellipse \(\mathcal E_\rho\) with radius $\rho > 1$  and continuous on
\(\mathcal{E}_\rho\). Further define the maximum of $g$ on the Bernstein ellipse, $M_{\rho,g}:=\max_{z\in\mathcal{E}_\rho} |g(z)|.$
The function $g(z)$admits a uniformly convergent Chebyshev polynomial expansion $g(z)=\sum_{n=0}^{\infty} t_nT_n(z)$ for $z\in[-1,1]$, and its Chebyshev coefficients satisfy $|t_0|\le M_{\rho,g}$ and $|t_n|\le 2M_{\rho,g}\rho^{-n}$ for all $n\ge1.$
Consequently, the truncated Chebyshev series $p_d(x):=\sum_{n=0}^{d}t_nT_n(x)$ satisfies $\|g-p_d\|_{\infty}\le\frac{2M_{\rho,g}}{\rho-1}\rho^{-d}.$ 
\end{theorem}

Naively one might think that the tightest bound for the convergence of Chebyshev interpolants can be derived by choosing the maximal possible $\rho$. For a function $g$ with singularities at $S = \{z_1, ..., z_d  \}$ this would specifically be $\rho_g:=\inf_{z_i\in S}[\max_{\pm} |z_i\pm\sqrt{z_i^2-1}|]$. However one can choose any $1<\rho<\rho_{g}$ to get a correct upper bound. The optimal choice of $\rho$ should ultimately balance the geometric decay of $\rho^{-d}$ against the non-decreasing $M_{\rho,g}$ on the Bernstein ellipse to obtain the tightest upper bound. Therefore, for a fixed degree $d$, one may select 
\begin{equation}\label{eq:optimal_rho}
\rho^\star_d =\operatorname*{arg\,min}_{1<\rho<\rho_{g}} \frac{M_{\rho,g}}{\rho-1}\rho^{-d}    
\end{equation}

\subsection{Super-Exponential Convergence and Exponential Sample Complexity}
Next we are going to show that with a very mild additional constraint on the growth of the target function we are capable of deriving a super-exponential upper bound on the convergence as well as exponential growth of the sample complexity.

\begin{theorem}[Super-exponential  convergence with exponential sample complexity]\label{theorem: super_algebraic_convergence and polynomial sample complexity}
     Assume a target function $g$ that grows not faster than super-exponentially as given by its maximum on a Bernstein ellipse of radius $\rho$, that is:
    \[
    M_{\rho,g}=\mathcal{O}\!\left(e^{C\rho^\alpha}\right),
    \]
    for some \(C,\alpha>0\) and for some positive integer $d$ the function $g(z)$ does not have any singularity inside a Bernstein ellipse with the radius $\left(\frac{d+1}{C\alpha}\right)^{\frac{1}{\alpha}}$ or in other words $\rho_g>\left(\frac{d+1}{C\alpha}\right)^{\frac{1}{\alpha}}$, then the $d-$degree Chebyshev interpolant approximating the target function  \(p(z)=\sum_{l=0}^{d} t_l T_l(z)\) converges super-exponentially, that is for any $c<\frac{1}{\alpha}$:
     \begin{equation}
     ||g-p||_{\infty} =  \mathcal{O}(d^{-cd})\, 
     \end{equation} 
     while its sample complexity is scaling exponentially, 
     \begin{equation}
     \Gamma(p) = \mathcal{O}(\exp(d))
     \end{equation}
\end{theorem}

\begin{proof}
    To simplify the calculation we are not directly searching for the optimal $\rho^\star$, c.f. Eq.~\ref{eq:optimal_rho}, but for the optimum with respect to a slightly adapted version of the optimality equation with the consideration of $\rho\gg 1$,
    \begin{equation}
        \frac{\partial}{\partial\rho} (\log(M_{\rho,g}\rho^{-(d+1)}))\stackrel{!}{=}0, \quad \Leftrightarrow \quad  C\alpha\rho^{\alpha-1}-
        \frac{d+1}{\rho}=0
    \end{equation}
    for large enough $d$ we may approximate the near optimal $\bar{\rho}$ simply with $\bar{\rho} = \left(\frac{d+1}{C\alpha}\right)^{\frac{1}{\alpha}}$. These approximations are all possible since we are still within the permissible range of radii for the Bernstein ellipse $1< \bar{\rho} < \rho_g$ and in the worst case there is a better $\rho^\star$ that leads to a  tighter bound. Inserting our $\bar{\rho}$ into the upper bound for the approximation error we get,
    \begin{equation}\label{eq: super_algebraic_convergence}
        \|g-h\|_{\infty} \leq\frac{2M_{\bar{\rho},g}}{\bar{\rho}-1}\bar{\rho}^{-d}  
         =\frac{M_0 \exp(\frac{d+1}{\alpha})}{\left(\frac{d+1}{C\alpha}\right)^{\frac{1}{\alpha}}-1} \left(\frac{d+1}{C\alpha}\right)^{-\frac{d}{\alpha}}
        =\mathcal{O}\left(d^{-c d}\right)
\qquad \text{for every } c < \frac{1}{\alpha}.
    \end{equation}
    where $M_0$ is some constant. For the proof of the exponential scaling of the the sample complexity upper bound we combine results from corollary~\ref{corr:Chebyshevl1} and Theorem~\ref{theorem: Bernstein_ellipse} to get,
    \begin{equation}\label{eq:exp_samp_compl}
    \Gamma(p)\leq(2+2\sqrt2)^d\| p\|_{T}\leq
    (2+2\sqrt{2})^d M_{\rho, g}[1+\frac{2(1-\rho^{-d})}{\rho-1}]<(2+2\sqrt{2})^d M_{\rho, g}\frac{\rho+1}{\rho-1}\,.
    \end{equation}
    From this formula we can deduce that a random, constant choice for $\rho$ within the permissible interval $1\leq \rho\leq \rho_g$ will lead to an exponentially scaling upper bound of the sample complexity. 
\end{proof}

With this we have extended the exponentially scaling upper bound for the sample complexity for truncated Jacobi-Anger series to all possible Chebyshev interpolants.
The basic insight that we use for the proof here is that we can employ two different optimal $\rho$'s. For the bound on the convergence we employ a polynomially with $d$ scaling $\rho$, while for the bound on the sample complexity we chose a constant $\rho$. 
We see that we get super-exponential convergence for very fast growing target functions $g$ as long as they do not have any singularities at finite distance from the origin of the complex plane. Asymptotically this statement is only true for entire target functions, or to be very specific, analytical continuations of the target function defined on the interval $[-1,,1]$ in the entire complex plane. This is due to a $\rho_{g}> \left(\frac{d+1}{C\alpha}\right)^{\frac{1}{\alpha}}$ that is growing with the degree of the approximation. Considering this constraint on the target functions, could we then simply take a logarithmic $d$ to get super-algebraic convergence as well as polynomial sample complexity? To realize that this is not as easy as it might seem at this point, we have to take a closer look at the kind of target functions we want to employ in a general quantum algorithmic framework. Ultimately, this is the reason why we decided to use $g$ as the function to interpolate in this section. $g$ will be $f/q$ in the following sections as we are going to show later that the target function of our QCPP $f$ is not the function we can approximate with logarithmic $d$. 

\subsection{Target Functions with Scaling Factor}
When we want to apply a function of a Hermitian operator on an arbitrary starting state $f(H)|\psi\rangle$ we typically want to guarantee that the target function $f$ is well approximated across the entire spectrum of the Hermitian operator $H$. In the main text we chose to normalize the Hermitian operator such that its spectrum is contained within the interval $[-1,1]$ and we did proceed with finding polynomial interpolants of $f$ for the interval $[-1,1]$. To be more transparent however one should then consider target functions $f(\cdot) \rightarrow f(\lambda \,\cdot)$ with an explicit scaling factor $\lambda$ that was chosen to ensure that the spectrum of $H/\lambda$ is contained in the interval $[-1,1]$. It is the general promise of quantum computing that its runtime scaling can be polynomial as a function of the problem size. The role of the problem size in our setting is played by the parameter $\lambda$ which therefore has to be given specific attention in our scaling investigations. In the case of real and imaginary time evolution the simulated time is another type of scaling factor that we also want to deal with separately. For example in the case of real time evolution there is the result of the no-fast forwarding theorem that states that the runtime scaling of a quantum algorithm can be and at the same should not be faster then linear in the simulated time. We subsume all scaling factors into one scaling factor $f(\cdot) \rightarrow f(\eta \, \cdot)$ and from now on search for polynomial interpolants  with a query as well as sample complexity that is polynomial in this scaling factor. As we can see however in the scaling equation for the sample complexity Eq.~\ref{eq:exp_samp_compl}, every scaling factor would lead to an at least  exponential scaling because of the presence of the maximum of the target function on the Bernstein ellipse at radius $\rho$, $M_{\rho,g}$ on the right side of the inequality. We therefore need to further demand that the maximum of the target function $g$ on the Bernstein ellipse is growing with respect to the scaling factor according to $M_{\rho,g} =  \mathcal{O}(\exp(C \rho^\alpha))$ for some $C>0$. This is not fulfilled for relevant target functions like, for example, real- $f(\cdot) = e^{i t \cdot}$ and imaginary- $f(\cdot) = e^{\tau (\cdot\, + 1)}$ time evolution. 

\subsection{Preconditioning and Refinement}
To separate the asymptotic behaviour with respect to the scaling parameter for the target functions of real- $f(z) = \exp(i t \lambda z) = \exp(i \eta_{\text{real}} z)$ and imaginary- $f(z) = \exp(-\tau \lambda (z + 1))= \exp(-\eta_{\text{imag}} (z + 1))$ time evolution we employ a preconditioning polynomial $q(z)$, that we choose to be sample complexity neutral $\Gamma(q) =1$. The goal is therefore to find a $q$ such that, there exists a positive constant $C>0$ and,
\begin{equation}\label{eq: super-algebraic-condition}
\max_{z \in \mathcal{E}_\rho}\left|\frac{f(z)}{q(z)}\right| = \mathcal{O}(\exp(C\rho^\alpha)) \,.
\end{equation}
such that $f/q$ satisfies the condition of Theorem~\ref{theorem: super_algebraic_convergence and polynomial sample complexity}. Remember here that both $f$ and $q$ do depend on the scaling parameter while the ratio is asymptotically constant with respect to the scaling factor. Ultimately we are going to be able to show that a precoditioning $q$ with a polynomially growing degree as a function of the scaling parameter is sufficient to get a maximum on the Bernstein ellipse that is asymptotically constant as a function of the scaling parameter.  
This would bring us in a position to define the interpolant as a product of polynomials, 
\begin{equation}
f = q h^\star \approx qh \quad \text{for} \quad h^\star = \frac{f}{q} = \sum\limits_{n=0}^\infty t_n T_n\,,
\end{equation}
with $h = \sum_{n=0}^{\mathcal{O}(log(d))} t_n T_n$ the Chebyshev refinement polynomial as a truncation of the infinite sum  $h^\star$ with logarithmic summands. 
What remains to be shown is the existence of such a sample complexity neutral $q$, that  accomplishes the separation of asymptotic behaviour and at the same time does enable exponentially small approximation error for an interpolant as defined above. This would ultimately mean that we also get polynomial sample complexity, $\Gamma(qh)= \Gamma(q)\Gamma(h) =  \Gamma(h) = \exp(\mathcal{O}(\log(d)))= \operatorname{poly}(d)$, as well as query complexity, $d + \mathcal{O}(\log(d)) = \mathcal{O}(d)$. Both results are consequences of Theorem~\ref{theorem: super_algebraic_convergence and polynomial sample complexity} for a $g = f/q$. 

We start by showing the capability of $q$ to make the scaling of the maximum on the Bernstein ellipse of radius $\rho$ asymptotically constant as a function of the scaling parameter.

Without loss of generality we may write $q$ in its factorised form such that, 
\begin{equation}\label{eq: protected_factor}
    q(z) = \prod\limits_{i=1}^d\frac{z - z_i}{|\Re[z_i]| + \sqrt{1+ \Im[z_i]^2}}\,.
\end{equation}
where the denominator is chosen to ensure the neutral sample complexity $\Gamma(q) = 1$. To make the following derivations palpable we introduce two reformulations of the task. First, we perform the Joukowski map $z=J(\omega)=\frac{1}{2}(\omega+\frac{1}{\omega})$ which is mapping the circumference of a circle of radius $\rho$ to the Bernstein ellipse of radius $\rho$ for $|\omega|>1$. This way we can examine the asymptotic scaling with respect to the magnitude of $\omega$ because $\max_{z\in\mathcal{E}_\rho}|f(z)/q(z)| \rightarrow \max_{|\omega| = \rho}|f(J(\omega))/q(J(\omega))|$, under the Joukowski map. Second, we are considering the logarithm of the ratio of $f$ and $q$ on the Bernstein ellipse, 
\begin{align}\label{eq:logf_vs_q}
    \log(\frac{f_{\text{real}}(J(\omega))}{q_{\text{real}}(J(\omega))})&=\sum\limits_{n=1}^{\infty}(i\delta_{n,1}  \frac{\eta_{\text{real}}}{2} +\sum\limits_{i}\frac{1}{n\omega_i^n})[\omega^n+\frac{1}{\omega^n}]+\sum\limits_i\log\left(\frac{|\Re[z_i]|+\sqrt{1+(\Im[z_i])^2}}{-\frac{\omega_i}{2}}\right)\\
    \log(\frac{f_{\text{imag}}(J(\omega))}{q_{\text{imag}}(J(\omega))})&=\sum\limits_{n=1}^{\infty}(-\delta_{n,1} \frac{\eta_{\text{imag}}}{2} +\sum\limits_{i}\frac{1}{n\omega_i^n})[\omega^n+\frac{1}{\omega^n}]+\sum\limits_i\log\left(\frac{|\Re[z_i]|+\sqrt{1+(\Im[z_i])^2}}{-\frac{\omega_i}{2}}\right)-\eta_{imag}\,,
\end{align}
where we have used the following expansion $\log(1-u)=-\sum_{n=1}^{\infty}\frac{u^n}{n}$ for $\log(q(z))$,
\begin{equation}\label{eq: logarithmic potential}
    \log(z-z_i)=\log[(-\frac{\omega_i}{2})(1-\frac{\omega}{\omega_i})(1-\frac{1}{\omega\omega_i})]=\log(-\frac{\omega_i}{2})-\sum_{n\geq1}^{\infty}\frac1n [\left(\frac{\omega}{\omega_i}\right)^n+\left(\frac{1}{\omega\omega_i}\right)^n]\,,
\end{equation}
with $z_i = \frac12(\omega_i + \frac{1}{\omega_i})$ the Joukowski map of the zeros of $q$. It is sufficient to choose a single unique root for $q$, $\omega_{\text{real}}$ or $\omega_{\text{imag}}$ respectively, to get rid of the scaling factor dependence in the infinite sums on the right hand side of the equations~\ref{eq: logarithmic potential}. We do leave the possibility however of having a multiplicity, $\kappa_{\text{real}}$ or $\kappa_{\text{imag}}$, of these roots,  
\begin{align}
 i \frac{\eta_{\text{real}}}{2} +\sum\limits_{i}\frac{1}{\omega_i} &=0 \iff \omega_{\text{real}} = 2 i \frac{\kappa_{\text{real}}}{\eta_{\text{real}}} \label{eq:omega_real}\\
-\frac{\eta_{\text{imag}}}{2} +\sum\limits_{i}\frac{1}{\omega_i} &= 0 \iff \omega_{\text{imag}} =  2 \frac{\kappa_{\text{imag}}}{\eta_{\text{imag}}}\label{eq:omega_imag}\,.
\end{align}
The remainder of the infinite sum after cancelling the first scaling parameter dependent term, does depend on the scaling parameter because it depends on the value of the root of $q$. We will show how the multiplicity of the root, $\kappa$, can now be used to make the remainder of the infinite sum asympotically constant with respect to the scaling parameter. The remainder of the infinite sums for real- and imaginary time are,  
\begin{align}\label{eq:rho_scaling}
\kappa_{\text{real}} \sum\limits_{n = 2}^\infty \frac{1}{n\omega_{\text{real}}^n}[\omega^n+\frac{1}{\omega^n}] & &
\kappa_{\text{imag}} \sum\limits_{n = 2}^\infty \frac{1}{n\omega_{\text{imag}}^n}[\omega^n+\frac{1}{\omega^n}]
\end{align}
The zeros of $q$ are the poles of $f/q$ for our cases of real and imaginary time evolution. As we know from Theorem~\ref{theorem: super_algebraic_convergence and polynomial sample complexity} the target function of the Chebyshev interpolant $h$, i.e. $f/q$ should be analytical within the Bernstein ellipse of radius $\rho = |\omega|$. This means we have to make sure later that $|\omega_i| > |\omega|$. For now we use this fact and  $|\omega|>1$ and $|\omega_i|>1$, thanks to the Joukowski map, to conclude that, somewhat counterintuitively, the dominating term in the remainder of the   infinite sum of the right hand side of equation~\ref{eq: logarithmic potential} is the one with the lowest order, $n=2$. Therefore, to make the infinite sum asymptotically constant, we can conclude that $\kappa$ has to scale according to $\kappa = \Omega(\eta^2)$ such that the sum does scale according to $\mathcal{O}(\rho^2)$ and is asymptotically constant with respect to the scaling parameter $\eta$. With this in mind, we put the solutions in Eq.~\ref{eq:omega_real} and Eq.~\ref{eq:omega_imag} through the Joukowski map back into Eq.~\ref{eq: protected_factor}, then we obtain the polynomials $q$ in Eq.~\ref{eq: real_protected_factor} and Eq.~\ref{eq: imag_protected_factor} in the main text in the limit of $\eta/\kappa\ll1$ with the choice of  $\kappa=d_q = \Omega(\eta^2)$.

To finish our argument on how the preconditioning polynomial $q$ accomplishes the separation of scaling with respect to the scaling parameter we need to argue that the rest of the terms of the right hand sides of equations~\ref{eq:logf_vs_q} have asymptotically constant real parts. An imaginary part for these terms would simply be an bounded oscillatory part of the function $f/q$ as we are considering here the logarithm of $|f/q|$, therefore it suffices to consider only the real part of the terms. The argumentation for real- and imaginary time evolution is slightly different here. For real time evolution  $\omega_{\text{real}}= 2i\kappa_{\text{real}} / \eta_{\text{real}}$, and therefore $\Im[z_{\text{real}}] = \frac{1}{2} (\frac{2\kappa_{\text{real}}}{\eta_{\text{real}}} - \frac{\eta_{\text{real}}}{2\kappa_{\text{real}}} )$ and $\Re [z_{\text{real}}]= 0$, we can reformulate, 
\begin{multline}
\kappa_{\text{real}}\log\left(\frac{|\Re[  z_{\text{real}}]|+\sqrt{1+(\Im[ z_{\text{real}}])^2}}{-\frac{\omega_{\text{real}}}{2}}\right) = \kappa_{\text{real}}\log\left(1+\left(\frac{\eta_{\text{real}}}{2\kappa_{\text{real}}}\right)^2 \right) +i\frac{3\pi\kappa_{\text{real}}}{2} = \\
= \frac{\eta_{\text{real}}^2}{4\kappa_{\text{real}}}+ \mathcal{O}(\frac{\eta_{\text{real}}^4}{\kappa_{\text{real}}^3}) +i\frac{3\pi\kappa_{\text{real}}}{2}\,,
\end{multline}
Since we have chosen the multiplicity to scale with the square of the scaling parameter, $\kappa_{\text{real}} = \Omega(\eta^2)$, we can conclude that the real part of the above terms is asymptotically constant with respect to the scaling parameter. For imaginary time evolution the terms on the right hand side of equation~\ref{eq:logf_vs_q} are,
\begin{multline}
\kappa_{\text{imag}}\log\left(\frac{|\Re[  z_{\text{imag}}]|+\sqrt{1+(\Im[ z_{\text{imag}}])^2}}{-\frac{\omega_{\text{imag}}}{2}}\right) - \eta_{\text{imag}} = 2\kappa_{\text{imag}}\log\left(1+\frac{\eta_{\text{imag}}}{2 \kappa_{\text{imag}}} \right) + i2\pi\kappa_{\text{imag}} - \eta_{\text{imag}}=\\ = - \frac{\eta_{\text{imag}}^2}{4\kappa_{\text{imag}}} + \mathcal{O}(\frac{\eta_{\text{imag}}^3}{\kappa_{\text{imag}}^2}) + i2\pi\kappa_{\text{imag}}\,.
\end{multline}
Since we have chosen the multiplicity of the root to scale quadratically in terms of the scaling parameter, we can conclude that also for imaginary time evolution the real part of the remainder on the right hand side of equation~\ref{eq:logf_vs_q} is asymptotically constant in terms of the scaling parameter. 

Finally we need to show that the preconditioning polynomial $q$ does not destroy the super-algebraic convergence of $p$ towards $f$, which should be inherited from the super-algebraic convergence of $h$ truncated at $d_h = \mathcal{O}(\log(\kappa))$ towards $h^\star$.  
To this end we prove that, on the interval $[-1, 1]$, $q(x)$ is bounded by $1$. For that we use the triangular inequality $\| u+v\|_2\leq \|u\|_2+\|v\|_2$. For $z \in [-1,1]$ we can therefore conclude, 
\begin{eqnarray}
    \begin{split}
        &u_i=(\Re [z_i], 0), \quad v_i=(z, \Im [z_i]), \quad \|(x, y)\|_2:=\sqrt{x^2+y^2}\\
        &|z-z_i|=\|(z-\Re [z_i], \Im [z_i])\|_2=\|v_i - u_i\|_2\leq \|u_i\|_2+\|v_i\|_2=|\Re [z_i]|+\sqrt{z^2+(\Im [z_i])^2}\\
        &|q(z)|=\prod_i\frac{|z-z_i|}{|\Re [z_i]|+\sqrt{1+(\Im [z_i])^2}}\leq\prod_m\frac{|\Re [z_i]|+\sqrt{z^2+(\Im [z_i])^2}}{|\Re [z_i]|+\sqrt{1+(\Im [z_i])^2}}\leq 1 \,.
    \end{split}
\end{eqnarray}
Based on Theorem~\ref{theorem: super_algebraic_convergence and polynomial sample complexity}, we further use the sub-multiplicativity of the supremum norm to get our final result,
\begin{equation}
    \|f-p\|_{\infty}=\|q(h^\star-h)\|_{\infty}\leq\|q\|_{\infty}\|h^\star-h\|_{\infty} = \mathcal{O}(\kappa^{-c\log\log(\kappa)})
\end{equation}
for any $0<c<\frac12$, because $\alpha=2$ according to the previous calculation.
From this, we can conclude that the interpolating polynomial $p = qh$ converges towards the target function $f$ super-algebraically.

\subsection{Query complexity of real- and imaginary time evolution}\label{appendix: section query complexity}
We finish the appendix section on the polynomial product interpolants with a formula that describes the asymptotic behaviour of the query complexity as a function of the approximation error, scaling parameter and sample complexity. We start by expressing the overall query complexity i.e. the degree $d$ of the product of the polynomials $p = qh$, as a sum, $d = \kappa + d_h$ with $\kappa$ the degree of $q$ and $d_h$ the degree of the refinement Chebyshev interpolant $h$. 

We start by further detailing the relationship between $\kappa$ and the scaling parameter $\eta$ with the help of the bound on the sample complexity, equation~\ref{eq:exp_samp_compl}, which we can further refine for the specific cases of real- and imaginary time evolution, 
\begin{equation}\label{eq: sample_complexity_upper_bound}
\Gamma(qh) = \Gamma(h) \leq (2+2\sqrt2)^{d_h}\exp(\frac{\eta^2}{8\kappa}\rho^2 + \mathcal{O}(\frac{\eta^3}{\kappa^2}\rho^3)) \frac{\rho + 1}{\rho-1}\,.
\end{equation}
in which we have calculated $M_{\rho, h}$ through the infinite sum in Eq.~\ref{eq:rho_scaling} obtained after the preconditioning, and $\Gamma_\star(p)$ is the upper bound of the sample complexity of the polynomial $p$.

To find a tight as possible and practical upper bound for the sample complexity, we retrace the steps of the proof of Theorem~\ref{theorem: super_algebraic_convergence and polynomial sample complexity} and search for the approximately optimal $\bar{\rho}$, this time taking into consideration also the dependence on the scaling parameter. Following these steps we get as the approximate derivative of the upper bound of the sample complexity 
\begin{equation}\label{eq: c21}
\frac{\eta^2}{4\kappa} \tilde\rho - \frac{2}{\tilde\rho^2 -1} =0 
\end{equation}
Assume for the remainder of the derivation that we choose the multiplicity of the root scaling with the square of the scaling parameter as defined above, however with a large prefactor such that we can further demand $\eta^2/\kappa \ll 1$. In this case, we can approximate the near optimal $\rho$ to be $\tilde{\rho} \approx (8\kappa/\eta^2)^{1/3}$. The dependence of the root multiplicity \(\kappa\) on the sample complexity \(\Gamma(p)\) will be determined after estimating the degree \(d_h\) of the Chebyshev interpolant \(h\). To express the degree of the Chebyshev refinement polynomial $h$ in terms of both the sample complexity $\Gamma(p)$ and the degree of the preconditioning polynomial $\kappa$, we follow the same strategy and try to find an approximate optimal bound for the approximation error with a special focus on the role of the scaling parameter, 
\begin{equation}
\epsilon = ||qh - f||_{\infty}\leq \frac{M_0 \exp(\frac{\eta^2}{8\kappa}\rho^2 + \mathcal{O}(\frac{\eta^3}{\kappa^2}\rho^3))}{\rho-1}\rho^{-d_h}\,,
\end{equation}
for some finite $M_0$. With the same procedure as detailed above in section~\ref{appendix section: Bernstein ellipse}, we get an approximately optimal Bernstein ellipse parameter of $\bar{\rho} = \sqrt{4\kappa d_h/\eta^2}$. The Bernstein ellipse parameter must remain
within the domain of analyticity of the
refinement target $h^\star=f/q$. Denoting by
$\rho_{h^\star}$ the ellipse parameter
corresponding to the nearest zero of $q$,
we therefore require $1<\bar{\rho}<\rho_{h^\star}.$ For the preconditioning polynomials considered
here, $\rho_{h^\star}\sim 2\kappa/\eta$ in the
regime $\kappa/\eta\gg 1$. Hence the above
choice of $\bar{\rho}$ is admissible when
$d_h<\kappa$ asymptotically. This condition
is satisfied by the logarithmic-refinement
construction $d_h=O(\log\kappa)$ for
sufficiently large $\kappa$. With this Bernstein ellipse parameter and large enough $d_h \gg 1$ we can derive the following upper bound for the degree of the Chebyshev refinement polynomial, 
\begin{equation}\label{eq: c23}
d_h =\mathcal{O}\left(\frac{\log(\frac{1}{\epsilon})}{W(\frac{8\kappa}{\eta^2}\log(\frac{1}{\epsilon}))}\right)\,,
\end{equation}
with the Lambert $W$ function. From the asymptotics of the Lambert $W$ function, $\lim\limits_{x\rightarrow\infty}W(x) = \log(x) + \mathcal{O}(\log\log(x))$, we can derive the asymptotic behaviour of the degree of the Chebyshev refinement polynomial, $d_h = \mathcal{O}(\log(1/\epsilon)/\log\log(1/\epsilon))$, which we may for clarity reasons rephrase to the weaker, and therefore still correct statement $d_h=\mathcal{O}(\log(1/\epsilon))$. Ultimately, we can conclude for the overall query complexity of QCPP for the product polynomials for real and imaginary time evolution.

The Bernstein ellipse parameters used to bound the sample complexity and approximation error may be chosen independently. For the sample complexity bound, the approximation \(\widetilde\rho\approx(8\kappa/\eta^2)^{1/3}\) requires \(\widetilde\rho\gg1\), or equivalently \(\eta^2/\kappa\ll1\). Indeed, $\log\frac{\rho+1}{\rho-1}=\frac{2}{\rho}+O(\rho^{-3}).$
At the proposed choice of \(\widetilde\rho\), the correction is \(O(\eta^2/\kappa)\), smaller than the leading contribution \(O((\eta^2/\kappa)^{1/3})\). The remainder in Eq.~\ref{eq: sample_complexity_upper_bound} is negligible relative to its quadratic term provided \(\eta\widetilde\rho/\kappa\ll1\). Substituting the optimized Bernstein ellipse parameter from Eq.~\ref{eq: c21} into Eq.~\ref{eq: sample_complexity_upper_bound}, we obtain, to leading order,
\begin{equation}\label{eq:C24}
\log\Gamma(qh)\lesssim d_h\log (2+2\sqrt2)+
3\left(\frac{\eta^2}{8\kappa}\right)^{1/3}.
\end{equation}
Provided that the contribution \(d_h\log(2+2\sqrt2)\) remains smaller than a fixed fraction of \(\log\Gamma_\star\), Eq.~\ref{eq:C24} yields \(\kappa=O(\eta^2/[\log\Gamma_\star]^3)\) within the admissible large-ellipse regime. Combining this with Eq.~\ref{eq: c23}, we obtain
\begin{equation}\label{eq: query_sample_complexity}
d = \kappa + d_h = \mathcal{O}(\frac{\eta^2}{[\log\Gamma_\star]^3}) + \mathcal{O}(\log(\frac{1}{\epsilon}))\,.
\end{equation}
Eq.~\ref{eq: query_sample_complexity} establishes the query-complexity scaling stated in Theorem~\ref{thm: query-sample} within the admissible asymptotic regime specified above. In particular, for prescribed approximation accuracy and sample complexity, the leading query complexity scales quadratically with the evolution parameter. This completes the proof of Theorem~\ref{thm: query-sample}.

We can conclude from this that we attain optimal, i.e. exponential scaling in terms of the approximation error $\epsilon$ and get a scaling with respect to the scaling parameter $\eta$ that is equivalent to trotter scaling if we are not willing to accept exponential scaling of the sample complexity $\Gamma(p)$. The derivation of Eq.~\ref{eq: query_sample_complexity} here also establishes the claimed result of Theorem~\ref{thm: query-sample} in the main text. With this we conclude the analytical investigation of the scaling behavior of QCPP and go on to test real world performance for numerically optimised interpolating polynomials in the next section.

In contrast to the general super-algebraic convergence established in the
previous section, the exponential convergence here is driven by the
\(\kappa^{-1}\)-suppression of the coefficient of \(\rho^\alpha\) in the
exponent of \(M_{\rho,g}\). Indeed, if
\(
M_{\rho,g}\sim \exp\!\left(\frac{C\rho^\alpha}{\kappa}\right),
\)
then increasing \(\kappa\) exponentially reduces the growth of
\(M_{\rho,g}\) on Bernstein ellipses. Through the Bernstein interpolation
estimate, this translates into a correspondingly sharper exponential
approximation-error bound.

\section{Numerical results}\label{app:numerics}
For our numerical results we first use a discrete cosine transform to get a Chebyshev interpolation polynomial that is exact $\frac{f(z_j)}{q(z_j)} = h(z_j)$, at the Chebyshev nodes $z_j = \cos( \frac{2j+1}{2d}\pi)$. 

We start with showing the query- and sample complexity tradeoff. For fixed $d$ and $d_q$ we calculated the Chebyshev interpolation polynomial as detailed above and derive numerically the sample complexity $\Gamma(p)=\Gamma(h)$ according to Eq.~\ref{eq: sampling_complexity}. We repeat this procedure for a large range of values of $d$, calculating for every $d$ the sample complexity for every possible $d_q \leq d$. We report here for every value of the  query complexity $d$ the minimal sample complexity over all possible choices of $d_q$, ignoring cases where we did not have an approximation error better than a threshold value $\epsilon$. The approximation error is evaluated by calculating the maximal difference of the polynomial and the target function over a dense grid on the interval $[-1, 1]$.

\begin{figure}[htp]
\centering
\includegraphics[width=0.96\linewidth]{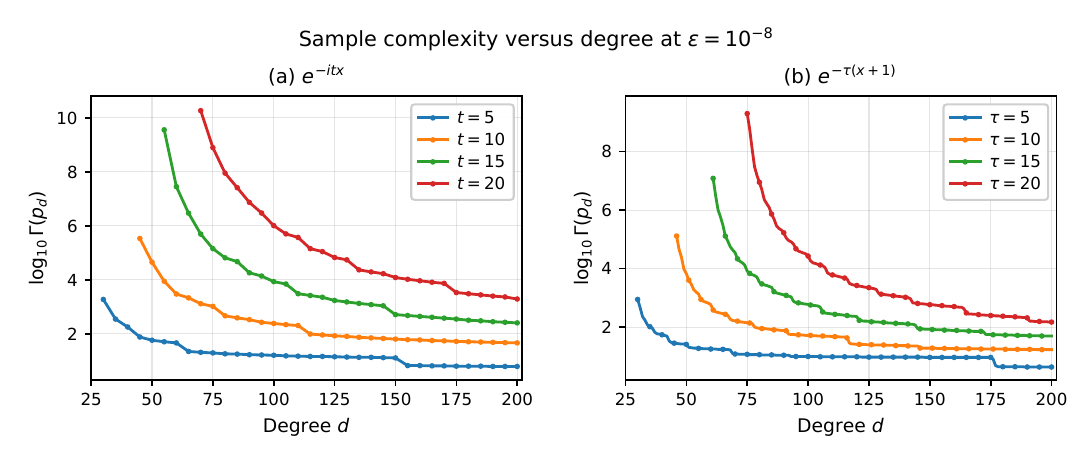}
\caption{The minimal sample complexity in log scale $\log\Gamma(p)$ in terms of the query complexity for implementing the $d-$degree polynomial $p$ approximating $f^{\textrm real}(x)=\exp(itx)$ and $f^{\textrm imag}(x)=\exp(-\tau(x+1))$.}
\label{fig:sample_query_tradoff}
\end{figure}

We have plotted in Fig.~\ref{fig:sample_query_tradoff} the minimal sample complexity $\Gamma(p)$ as a function of the query complexity $d$ for the polynomials interpolating real and imaginary time evolution with a fixed threshold approximation error $\epsilon = 10^{-8}$. All the curves start with the optimal query complexity, below which no lower degree polynomial attains the given threshold approximation error. The sample complexity decreases with increasing query complexity for a fixed threshold approximation error $\epsilon$. Since the argument of the exponential functions in both cases of real and imaginary time evolution does only depend linearly on simulated real or imaginary time there is only one distinct root of the polynomial $q$. Therefore a decrease in the sample complexity is due solely to an increase of the multiplicity of this specific root. Intuitively we can say that the sample complexity neutral function $q$ is capable of a better approximation of the target function such that the Chebyshev interpolating polynomial $h$ does not have to be of a particularly high degree which ultimately reduces the sample complexity.  

\begin{figure}[htp]
\centering
\includegraphics[width=0.96\linewidth]{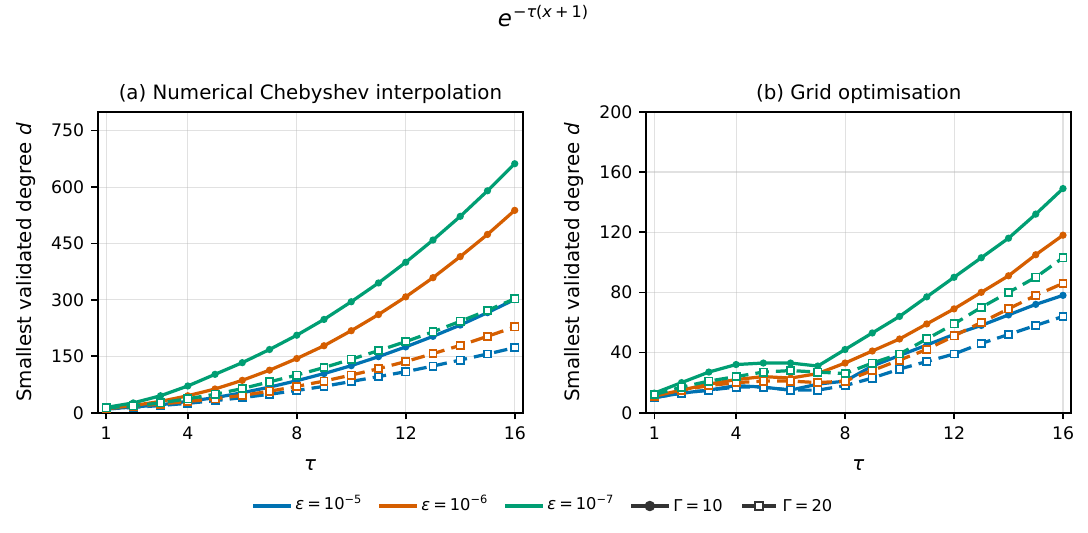}
\caption{
\textbf{Imaginary time evolution.}
Smallest validated polynomial degree \(d\) required to approximate
\(f^{\text{imag}}=e^{-\tau(x+1)}\) uniformly on \([-1,1]\), subject to
\(\lVert p-f^{\text{imag}}\rVert_\infty\leq\epsilon\) and the weighted
root-product constraint \(\Gamma(p)\leq\Gamma\).
Panel~(a) shows the results obtained by numerical Chebyshev interpolation,
whereas panel~(b) shows the results obtained by grid optimization.
Colours indicate the approximation error tolerance \(\epsilon\).
Solid curves with filled circles correspond to \(\Gamma=10\), while dashed
curves with open squares correspond to \(\Gamma=20\).
}
\label{fig:imaginary-exponential-degree}
\end{figure}

\begin{figure}[htp]
\centering
\includegraphics[width=0.96\linewidth]{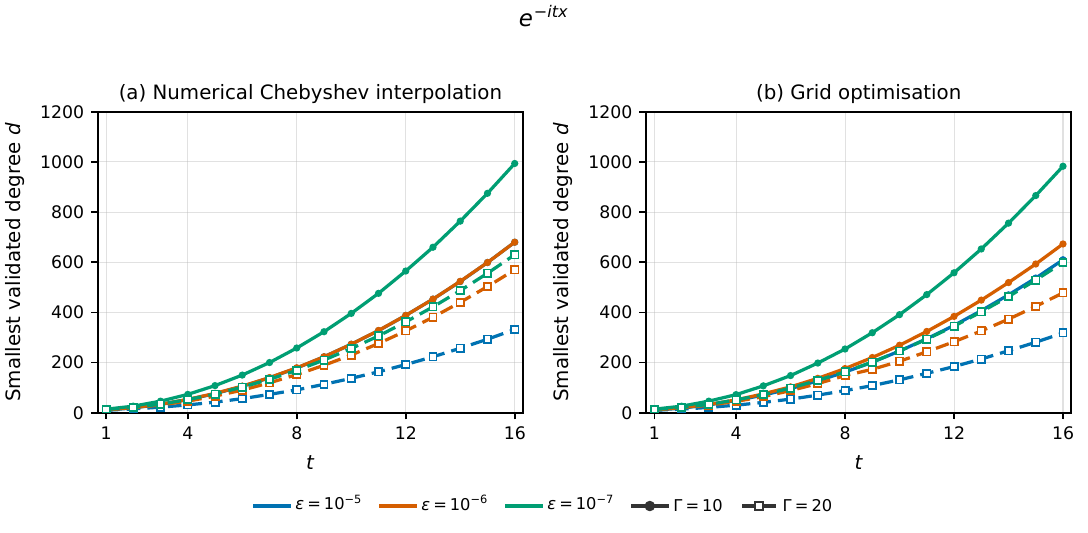}
\caption{
\textbf{Real time evolution.}
Smallest validated polynomial degree \(d\) required to approximate
\(f^{\text{real}}(x)=e^{-itx}\) uniformly on \([-1,1]\), subject to
\(\lVert p-f^{\text{real}}\rVert_\infty\leq\epsilon\) and the weighted
root-product constraint \(\Gamma(p)\leq\Gamma\).
Panel~(a) shows the results obtained by numerical Chebyshev interpolation,
whereas panel~(b) shows the results obtained by grid optimization.
Colours indicate the approximation error tolerance \(\epsilon\).
Solid curves with filled circles correspond to \(\Gamma=10\), while dashed
curves with open squares correspond to \(\Gamma=20\).
All reported approximants were verified on a dense validation grid using
the modulus of the complex approximation error. The grid optimization performs slightly better than the numerical Chebyshev interpolation, which reduces the query complexity by  $12.4$ on average.
}
\label{fig:real-exponential-degree}\end{figure}

Next we numerically investigate the scaling of the query complexity as a function of simulated real and imaginary time. To this end, we calculate the sample complexity for fixed real or imaginary time. These calculations are executed up to a query complexity $d$ such that the minimal sample complexity as well as the approximation error, with respect to all possible $d_q\leq d$ are both below their threshold values $\epsilon$ and $\Gamma_\star$. The sample complexity $\Gamma(h)$ and the approximation error are numerically evaluated in the same way as above. We have plotted in Fig.~\ref{fig:imaginary-exponential-degree}(a) and Fig.~\ref{fig:real-exponential-degree}(a) the minimal query complexity $d$ in terms of the simulated time $t$ and $\tau$ respectively. Increasing the sample complexity threshold $\Gamma_\star$ relaxes the query complexity constraint and therefore generally lowers the first feasible $d_{\text{min}}$. The numerical calculations confirm the findings of the analytical treatment in Eq.~\ref{eq: query_sample_complexity}: for fixed threshold approximation error and fixed threshold sample complexity, the optimal query complexity follows a quadratic scaling law as a function of $t$ and $\tau$, c.f. Fig.~\ref{fig:imaginary-exponential-degree}(a) and Fig.~\ref{fig:real-exponential-degree}(a). We obtain the following empirical relations, fitting the numerical results:
\begin{eqnarray}\label{eq: query_numerical_extrapolation}
\begin{split}
& d^{\mathrm{real}}\approx36.90\frac{t^2}{(\log \Gamma)^3}+4.12\log(\frac{1}{\epsilon}),  \qquad d^{\mathrm{imag}}\approx22.26\frac{\tau^2}{(\log \Gamma)^3}+2.35\log(\frac{1}{\epsilon})
\end{split}
\end{eqnarray}
These fits are numerical extrapolations over the finite sample complexity budget range of $\Gamma=10, 30, 60, 90$ and the approximation error $\epsilon=10^{-5}, 10^{-6}, 10^{-7}$, where we have used the functional form of the inverse logarithm of the sample complexity derived in Eq.~\ref{eq: query_sample_complexity}. The numerical fitting reaches an $R\sim 0.93$, which corroborates qualitatively the above mentioned analytical relations.

\begin{table}[t]
\centering
\caption{Empirical quadratic scaling of query complexity extrapolated from Eq.~\ref{eq: query_numerical_extrapolation}.}
\label{tab:degree-scaling}
\renewcommand{\arraystretch}{1.35}
\setlength{\tabcolsep}{10pt}
\begin{tabular}{cccc}
\toprule
Error tolerance
& Trotter query complexity
& $\Gamma=10$
& $\Gamma=20$
\\
\midrule
$\epsilon=10^{-3}$
&
$10^{3}t^{2}$
&
$\begin{aligned}
d_{\mathrm{real}} &= 3.02\,t^{2}+28.46\\
d_{\mathrm{imag}} &= 1.82\,\tau^{2}+16.23
\end{aligned}$
&
$\begin{aligned}
d_{\mathrm{real}} &=  1.37\,t^{2}+28.46\\
d_{\mathrm{imag}} &=  0.83\,\tau^{2}+16.23
\end{aligned}$
\\[1.2em]
\midrule
$\epsilon=10^{-4}$
&
$10^{4}t^{2}$
&
$\begin{aligned}
d_{\mathrm{real}} &= 3.02\,t^{2}+37.95\\
d_{\mathrm{imag}} &= 1.82\,\tau^{2}+21.64
\end{aligned}$
&
$\begin{aligned}
d_{\mathrm{real}} &= 1.37\,t^{2}+37.95\\
d_{\mathrm{imag}} &= 0.83\,\tau^{2}+21.64
\end{aligned}$
\\
\bottomrule
\end{tabular}
\end{table}

We have compared the query complexity of QCPP for real and imaginary time evolution with first order Trotterisation at different residual errors and sample complexity budgets. They are presented in Table~\ref{tab:degree-scaling}.  The exponential approximation error convergence offers a better query complexity scaling for implementing the real and imaginary time evolution functions compared to Trotter, which converges only polynomially in terms of residual error.

\subsection{Further Optimizing the Sample Complexity}

In the previous section we chose the technique of discrete cosine transform to get a linear superposition of Chebyshev polynomials with Chebyshev coefficients $t_m$. The sample complexity was a direct consequence of this choice. However it is also possible to rephrase the process of finding Chebyshev coefficients: find a linear superposition of Chebyshev coefficents such that the sample complexity is minimal under the constraint that the approximation error is not larger than some threshold value $\epsilon$. The sample complexity is a quantity that depends implicitly on the Chebyshev coefficients however in corollary~\ref{corr:Chebyshevl1} we prove an upper bound defined by the $l_1$ norm of the Chebyshev coefficients which can serve here as a proxy for our minimisation procedure:
\begin{equation}
\{t_m\}_{m=0}^{d_h} = \arg \min \sum\limits_{l=0}^{d_h} |t_m|\,,
\end{equation}
subject to,
\begin{equation}\label{eq: error_constraints}
\left|f(x_j)-q(x_j)\sum_{m=0}^{d_h}T_m(x_j)t_m\right|\leq\epsilon,        \qquad\forall x_j\in\mathcal{G} \,,
\end{equation}
for a suitably dense grid of points, $\mathcal{G}$, of the interval $[-1,1]$. With a surrogate vector $s:=(s_0,\ldots,s_{d_h})^{\mathsf T}\geq 0,$ that is there to represent the absolute value of the Chebyshev coefficients we are able to rephrase the minimisation procedure of Eq.~\ref{eq: error_constraints} into a  linear program:
\begin{equation}
\boxed{
\begin{aligned}
\operatorname{minimize}
\quad&
\mathbf{1}^{\mathsf T}s
\\
\text{subject to}\quad&
-\epsilon\mathbf{1}
\leq \Re(f-At)
\leq\epsilon\mathbf{1},
\\&
-\epsilon\mathbf{1}
\leq \Im(f-At)
\leq\epsilon\mathbf{1},
\\
&
-s_m\leq \Re(t_m)\leq s_m\qquad -s_m\leq \Im(t_m)\leq s_m\qquad 0\leq m\leq d_h.
\end{aligned}
}    
\end{equation}
with vectors $A$, $f$ and $h$ defined such that, 
\begin{align}
A_{jm}:=q(x_j)T_m(x_j),
\qquad t:=(t_0,\ldots,t_{d_h})^{\mathsf T}\qquad
f_j:=f(x_j)
\qquad \forall x_j \in \mathcal{G}
\end{align}
for complex-valued target functions, the linear program is implemented by separating the real and imaginary components of the coefficients and residuals. The resulting polynomial is subsequently validated using the complex modulus of the approximation error on a dense grid, and its sampling-normalization overhead \(\Gamma(p)\) is evaluated directly from its polynomial roots.

Subsequently, we are numerically optimizing a Chebyshev interpolation polynomial with the goal of minimizing the sample complexity while having a bound on the approximation error as a constraint for a linear program. We are reporting as above the minimal query complexity $d$ such that any of all the possible choices of $d_h\leq d$ provides us with some result that has a sample complexity which is smaller than a chosen threshold sample complexity. The results of this procedure are shown in Fig.~\ref{fig:imaginary-exponential-degree} and Fig.~\ref{fig:real-exponential-degree} panels $b)$ respectively. As expected this procedure does provide us with further reduced query complexity albeit with considerably higher success for the case of imaginary time evolution. The query complexity graphs also reveal some numerical issues that result in a non-monotonic increase of the query complexity as a function of simulated time.

\end{document}